\documentclass{birkjour}
\usepackage{graphicx}
\usepackage{amssymb,latexsym}
\usepackage{amsmath}
\usepackage{epsf}
\usepackage{pdfsync}
\usepackage[british]{babel}
\usepackage{epsfig}
\usepackage[parfill]{parskip}
\usepackage[matrix,arrow,color]{xy}
\usepackage[usenames]{color}
\usepackage{hyperref}

\input{xy}
\xyoption{all}

\input{epsf}

\numberwithin{equation}{section}
\catcode`@=12

\newcommand{\ba}{\begin{eqnarray*}}
\newcommand{\ea}{\end{eqnarray*}}
\newcommand{\ban}{\begin{eqnarray}}
\newcommand{\ean}{\end{eqnarray}}

\newcommand{\Tr}{{\rm Tr\,}}

\newcommand{\gr}{{\rm gr}}

\newcommand{\IZ}{\mathbb{Z}}
\newcommand{\IC}{\mathbb{C}}

\newcommand{\IN}{\mathbb{N}}

\newcommand{\frM}{\frak{M}}
\newcommand{\frN}{\frak{N}}
\newcommand{\frZ}{\frak{Z}}
\newcommand{\frS}{\frak{S}}
\newcommand{\frF}{\frak{F}}

\renewcommand{\Im}{\ensuremath{\mathfrak{Im}}}
\renewcommand{\Re}{\ensuremath{\mathfrak{Re}}}

\newcommand{\cW}{{\mathcal W}}
\newcommand{\cN}{{\mathcal N}}
\newcommand{\cM}{{\mathcal M}}
\newcommand{\cS}{{\mathcal S}}
\newcommand{\cB}{{\mathcal B}}

\newcommand{\cH}{{\mathcal H}}

\newcommand{\cE}{{\mathcal E}}
\newcommand{\cO}{{\mathcal O}}
\newcommand{\cC}{{\mathcal C}}

\newcommand{\cR}{{\mathcal R}}
\newcommand{\cZ}{{\mathcal Z}}

\newcommand{\cK}{{\mathcal K}}

\newcommand{\cV}{{\mathcal V}}
\newcommand{\cY}{{\mathcal Y}}

\newcommand{\sfA}{{\mathsf{A}}}
\newcommand{\sfe}{{\mathsf{e}}}

\newcommand{\sfZ}{{\mathsf{Z}}}
\newcommand{\sfQ}{{\mathsf{Q}}}
\newcommand{\sfD}{{\mathsf{D}}}

\newcommand{\sfR}{{\mathsf{R}}}
\newcommand{\sfX}{{\mathsf{X}}}
\newcommand{\sfM}{{\mathsf{M}}}
\newcommand{\sfC}{{\mathsf{C}}}

\newcommand{\mut}{{\mathsf{mut}}}

\newcommand{\NDT}{{\tt NC}}
\newcommand{\DT}{{\tt DT}}
\newcommand{\BPS}{{\tt BPS}}

\newcommand{\mbf}[1]{{\boldsymbol {#1} }}
\newcommand{\complex}{{\mathbb C}} 
\newcommand{\zed}{{\mathbb Z}} 
\newcommand{\real}{{\mathbb R}} 
\newcommand{\torus}{{\mathbb T}}

\def\e{{\,\rm e}\,}

\newcommand{\ch}{{\rm ch}}
\newcommand{\tch}{{\widetilde{\rm ch}}}
\newcommand{\Ch}{{\rm Char}}

\def\ii{{\,{\rm i}\,}}
\def\dd{{\rm d}}

\newcommand{\Hom}{\mathrm{Hom}}

\newcommand{\motive}{\mathbb{L}}

\newcommand{\Gammaw}{{\widehat\Gamma}}

\def\beq{\begin{equation}}
\def\bee{\begin{equation}}
\def\eeq{\end{equation}}
\def\bea{\begin{eqnarray}}
\def\eea{\end{eqnarray}}
\def\bd{\begin{displaymath}}
\def\ed{\end{displaymath}}

\newcommand{\Cint}{\int\kern-10.5pt-\kern7pt}

\newcommand{\PP}{{\mathbb{P}}}

\newcommand{\be}{\begin{equation}}
\newcommand{\ee}{\end{equation}}

\newcommand\fverbit{\egroup\item[\fbox{\unhbox\pippobox}]}
\newbox\pippobox

\renewcommand{\thesection}{\arabic{section}}
\renewcommand{\theequation}{\thesection.\arabic{equation}}

\def\a{\alpha}

\def\b{\beta}

{

\def\eps{\epsilon}

\def\L{\Lambda}
\def\O{\omega}

\def\be{\begin{equation}}
\def\ee{\end{equation}}
\def\bea{\begin{eqnarray}}
\def\eea{\end{eqnarray}}

\begin{document}

%
%
%

\title[Instanton counting and wall-crossing for orbifold quivers]{Instanton counting and wall-crossing for \\ orbifold quivers}


\author[M. Cirafici]{Michele Cirafici}

\address{Centro de An\'{a}lise Matem\'{a}tica, Geometria e Sistemas
Din\^{a}micos\\ Departamento de Matem\'atica\\ Instituto Superior T\'ecnico\\
Av. Rovisco Pais, 1049-001 Lisboa, Portugal}

\email{cirafici@math.ist.utl.pt}

\author[A. Sinkovics]{Annamaria Sinkovics}

\address{Department of Applied Mathematics and
  Theoretical Physics
\\ Centre for Mathematical Sciences, University of
Cambridge \\ Wilberforce Road, Cambridge CB3 0WA, UK}

\email{A.Sinkovics@damtp.cam.ac.uk}

\author[R.J. Szabo]{Richard~J.~Szabo}

\address{Department of Mathematics\\ Heriot--Watt
  University\\
Colin Maclaurin Building, Riccarton, Edinburgh EH14 4AS, UK \\ and\\ 
Maxwell Institute for Mathematical Sciences, Edinburgh, UK}

\email{R.J.Szabo@hw.ac.uk}




\begin{abstract}
Noncommutative Donaldson--Thomas invariants for abelian orbifold singularities can be studied via the enumeration of instanton solutions in a
six-dimensional noncommutative $\cN=2$ gauge theory; this construction is based on the
generalized McKay correspondence and identifies the instanton
counting with the counting of framed representations of a quiver which is naturally associated to the geometry of the
singularity. We extend these constructions to compute BPS
partition functions for higher-rank refined and motivic noncommutative Donaldson--Thomas
invariants in the Coulomb branch in terms of gauge theory variables and orbifold data. We
introduce the notion of virtual instanton quiver associated with the
natural symplectic
charge lattice which governs the quantum wall-crossing behaviour of BPS states in this context. The McKay correspondence naturally connects our
formalism with other approaches to wall-crossing based on quantum monodromy operators and cluster algebras.
\end{abstract}

\maketitle




\section{BPS states on local threefolds}

The spectrum of BPS states in supersymmetric string compactifications has a
subtle dependence on the background moduli
\cite{Cecotti:1992rm,Seiberg:1994rs,Denef:2007vg}; in the following we
will deal exclusively with the type~IIA string theory. As these moduli are
varied physical states can decay into fundamental constituents or form
bound states. As a result the single particle Hilbert space can lose
or gain a factor. BPS states can be realized by an appropriate
configuration of D-branes wrapping calibrated cycles in the internal
geometry which are mathematically characterized by certain enumerative
invariants. These invariants must also behave according to the
same pattern; this is the content of the theory of generalized
Donaldson--Thomas invariants \cite{joycesong,Kontsevich:2009xt}. On
crossing a wall of marginal stability, where physical states decay or
bound states are formed, the generalized Donaldson--Thomas invariants
change according to a wall-crossing formula. The matching between
mathematical predictions and physical expectations has been the focus
of much recent activity, see e.g.~\cite{Gaiotto:2008cd,Dimofte:2009bv,Gaiotto:2009hg,Cecotti:2009uf,Dimofte:2009tm,Gaiotto:2010be}.

It is however difficult to translate this picture into practice since
for a generic Calabi--Yau threefold the computation of
Donaldson--Thomas invariants is a rather daunting task. Thus there is
a need for having a controlled setup where conjectures can be stated
explicitly and computational evidence can be provided. Toric
Calabi--Yau manifolds are a commonly used playground since the torus
action allows for the use of powerful localization formulas to compute explicitly the enumerative invariants and therefore the spectrum of physical states.

For local toric Calabi--Yau threefolds the situation is as
follows \cite{szendroi,Ooguri:2008yb,Aganagic:2009kf}. The moduli space of
vacua is divided into chambers by walls of marginal stability in
codimension one. In one
of these chambers, corresponding to the large radius classical geometry, the BPS enumerative invariants are computed directly
by topological string theory and coincide with the ordinary
Donaldson--Thomas invariants, or equivalently with other enumerative
invariants such as the Gromov--Witten and Gopakumar--Vafa
invariants. The other chambers are in principle accessible from this
one by using wall-crossing formulas. In the chamber relevant to this paper the
enumerative invariants are captured by algebraic structures encoded in
a quiver. This chamber is non-geometric in the sense that it describes a
region in the moduli space of vacua where conventional geometry breaks
down due to quantum effects. Heuristically this region can be thought
of as arising in the limit where one or more cycles in the Calabi--Yau
manifold shrink to zero size but their quantum volume, as measured by
the $B$-field or any Ramond--Ramond field (depending on the
compactification), is still non-zero. The chamber in which this happens,
typically near conifold or orbifold points in the vacuum moduli space,
will be refered to as the noncommutative crepant resolution chamber.

The purpose of this paper is to investigate this chamber from the
worldvolume point of view of the D-brane configurations. From this
perspective the Donaldson--Thomas invariants count (generalized) instanton
solutions in the D6-brane worldvolume gauge theory which
characterize how many lower-dimensional branes are stably bound to the
D6-branes. The problem reduces to the classification and enumeration
of these instanton solutions. In the ordinary Donaldson--Thomas
chamber this was done in
\cite{qfoam,Jafferis:2007sg,Cirafici:2008sn}. This formalism was
adapted in \cite{Cirafici:2010bd} to study noncommutative
Donaldson--Thomas invariants associated with toric orbifold
singularities. A key ingredient in this approach is the generalized
McKay correspondence; it implies
that the noncommutative Donaldson--Thomas invariants can be computed
via a certain quantum mechanics based on a quiver diagram
associated with the singularity. This quiver matrix quantum mechanics
provides an efficient tool to compute explicitly the BPS spectrum of
D-brane bound states in terms of combinatorial data, namely coloured
partitions. 
This approach allows an explicit computation of higher rank invariants in the Coulomb branch, that is when the gauge group is broken down to $U(1)^N$. This phase turns out to be very rich and to depend on the specific boundary conditions chosen for the gauge field at infinity. By considering arbitrary boundary condition in each $U(1)$ factor one finds a rather intricate enumerative problem. Unfortunately there does not seem to be any straightforward extension of the formalism to capture genuinely nonabelian instanton configuration. This is a pretty much open problem and the only partial results in the literature \cite{stoppa,toda} hold for the simplest $\complex^3$ geometry. In this lucky case the walls of marginal stability corresponding to nonabelian bound states can be reached directly by tuning the B--field without crossing any other wall, and thus the wall--crossing formula can be applied directly to compute nonabelian higher rank BPS degeneracies.

The quiver matrix model comes with a very explicit dictionary, developed in \cite{Cirafici:2010bd}, between algebraic and geometric quantities. 
We will see that this dictionary provides a very powerful tool to
investigate structures associated with BPS states via the McKay
correspondence. Following the development in
\cite{Cecotti:2010fi}, a good strategy to approach the enumerative
problem of BPS states is to study objects that are \textit{invariant}
as one moves around the moduli space. Examples of such objects were
found in~\cite{Cecotti:1992rm} and more recently by Kontsevich and
Soibelman \cite{Kontsevich:2009xt} who showed that the condition for
a certain product of operators to be invariant on the moduli space is
equivalent to a wall-crossing formula for the degeneracies of BPS
states. To the BPS states which are captured by quivers we can
associate another algebraic structure that is (to some extent) invariant, called a
quantum cluster algebra. This is constructed by applying a sequence of
mutations to the quiver. Each mutation is the quantum mechanics
equivalent of a Seiberg duality and determines a jump in the BPS
spectrum, crossing a wall of the second
kind~\cite{Kontsevich:2009xt,Aganagic:2010qr}. The cluster algebra
encompasses all possible mutations. We show that to each orbifold
singularity of the class studied in~\cite{Cirafici:2010bd} it is
possible to canonically associate a quiver model, which we call a
virtual quiver, from which a cluster algebra can be
defined. Physically this can be achieved by giving ``enough'' masses
to bifundamental strings stretched between D-branes. We will find that
the structure of this virtual quiver, as well as the associated
quantum algebras, is completely determined by the McKay
correspondence. This opens up exciting possibilities for the study of
BPS states. Recall that in the four-dimensional case for du Val
singularities the McKay correspondence implies an action of an affine
Kac--Moody algebra on the instanton moduli space. Because of this
action the partition function of $\cN=4$ supersymmetric Yang--Mills
theory on the resolution of the singularity with fixed boundary
condition at infinity is a character of the affine Kac--Moody algebra. The analogy here with the quantum cluster algebra characterizing the BPS spectrum is rather tantalizing and deserves further investigation.

Another application we will describe is the use of instanton quivers to
set up the computation of motivic Donaldson--Thomas invariants in the
noncommutative crepant resolution chamber. Motivic Donaldson--Thomas invariants were
introduced by Kontsevich and Soibelman \cite{Kontsevich:2009xt}, who
used them to prove and conjecture wall-crossing identities for the ``numerical''
Donaldson--Thomas invariants. They represent a sort of
categorification of the Donaldson--Thomas invariants. Physically this means that the motivic invariants capture the
homological algebra of the Hilbert space of BPS states; they
represent regions of the moduli space of vacuum states defined via the
locus of relations associated with the instanton quiver. There is an
underlying K-theory of varieties (or better of stacks) whose
generator is a certain parameter $\motive^{\frac12}$. An integration map
connects this K-theory with ordinary refined invariants in the sense of
\cite{Iqbal:2007ii}, and with the ordinary numerical invariants. This
integration map should be thought of as the mathematical analog of
taking the (refined) Witten index over the Hilbert space of BPS
states; the inverse process, which associates to a number a certain
space such that the number is an invariant of that space,
is precisely the key point of the whole categorification process. These
issues were investigated in~\cite{Dimofte:2009bv,Dimofte:2009tm}
providing substantial evidence for the conjectured correspondence between motivic and
refined invariants. We may regard the motivic theory as a step towards
providing the geometric category pertinent to the topological
quantum field
theory underlying the D6-brane worldvolume gauge theory.

This paper is organized as follows. In sections 2--6 we set up the problem and summarize the results of our long
paper~\cite{Cirafici:2010bd} in a somewhat informal way that we hope
is more accessible to a wider audience. In particular we introduce what we call \textit{stacky gauge theories} which naturally compute classes of BPS invariants labelled by gauge theoretical boundary conditions. Then we put our formalism at work in Section 7 which includes our first new result, a partition function of refined BPS invariants. Precisely as in~\cite{Cirafici:2010bd}, the stacky gauge theory predicts that the partition function counts states weighted by the instanton action, contrary to other approaches which exist in the literature. Section 8 deals with physical D--brane charges. We find that our approach naturally endows the lattice of fractional brane charges with the K--theory intersection pairing. The relevant quiver is therefore not the ordinary McKay quiver but what we call a \textit{virtual instanton quiver}. We conjecture that such a quiver can always be reached by a superpotential deformation which gives masses to bi--fundamental strings. In Section 9 we adapt a formalism to compute motivic invariant to our instanton quivers and extend it to deal with generic Coulomb branch invariants for an arbitrary boundary condition. Finally in Sections 10 and 11 we study the connection between our formalism and other new approaches
to wall-crossing such as the theory of cluster
algebras, the quantum monodromy theory of \cite{Cecotti:2010fi}, and the
motivic Donaldson--Thomas theory. Our main goal here is to simply build a
bridge between these concepts; their connections deserve to be investigated further. In particular we suggest that the correct object to consider is the virtual instanton quiver. Once this step is taken, every ingredient is rephrased in terms of representation theory data. The stacky gauge theory approach allows to compute the central charges by using the McKay correspondence to fix their moduli dependence. In particular the wall--crossing formula and the cluster algebra formalism appear to depend on the boundary conditions. This approach suggests that the Seiberg--like dualities of the quiver quantum mechanics should be properly studied in the context of virtual quivers.

\section{Quivers and noncommutative crepant resolutions}

The chamber of the moduli space of string vacua that we consider is ``non-geometric'' in nature. By this we mean that its target space description does not correspond to ordinary geometry but requires more abstract tools. A recurrent theme in modern mathematics is that when ordinary geometry is not apt to describe a space, a perfectly good alternative (and sometimes conceptually deeper) description can be given in terms of algebraic structures defined over the space. Such is the case for example in noncommutative geometry where appropriate algebras of functions or operators over a space are used to characterize the space {\it itself} and its geometry. 

Many local Calabi--Yau threefolds admit a similar description in terms
of the representation theory of a quiver. A quiver
$\sfQ=(\sfQ_0,\sfQ_1)$ is an algebraic entity defined by a set of
nodes $\sfQ_0$ and by a set of arrows $\sfQ_1$ connecting the
nodes. To the arrows one might associate a set of relations
$\sfR$. The \textit{path algebra} of the quiver is defined as the
algebra of all possible paths in the quiver modulo the ideal generated
by the relations; the product in the algebra is the concatenation of
paths whenever this makes sense and $0$ otherwise. This algebra will
be denoted as $\sfA = \complex \sfQ / \langle\sfR \rangle$. A
representation of the quiver $\sfQ$ can be constructed by associating
a complex vector space to each node and a linear map between vector
spaces for each arrow, respecting the relations $\sfR$. When
appropriate conditions are met the moduli space of representations of
the quiver $\sfQ$, where all vector spaces have dimension one, is a
smooth toric Calabi--Yau variety $X$; this is the crepant resolution
of an abelian orbifold singularity $\complex^3 / \Gamma$ provided by
the $\Gamma$-Hilbert scheme $\mathrm{Hilb}^{\Gamma}
(\complex^3)$. Below we will regard this resolved geometry as
describing a certain ``large radius phase'' of D-branes on
the singularity.

Under certain circumstances the path algebra $\sfA$ associated with the quiver $\sfQ$ is itself a crepant resolution of the abelian singularity $\complex^3 / \Gamma$, called the noncommutative crepant resolution. In this case one replaces the singular space with an algebra whose center is the coordinate algebra of the singularity. In this sense $\sfA$ is a desingularization of $\complex^3 / \Gamma$. One can furthermore prove that $\sfA$ enjoys many nice properties expected from a crepant resolution.

For the abelian orbifold singularities $\complex^3 / \Gamma$ we wish to consider, a natural quiver is provided by the McKay quiver. This is defined in terms of the representation theory data of $\Gamma$. The vertex set $\sfQ_0$ is identified with the group $\widehat\Gamma$ of irreducible (one-dimensional) representations $\rho_r$ of $\Gamma$; the trivial representation is denoted $\rho_0$. The number of arrows going from node $s$ to node $r$ is $a_{sr}^{(1)}$, where in general $a_{sr}^{(i)}$ is defined as the multiplicities in the tensor product decomposition
\begin{equation} \label{tensordecomp}
\mbox{$\bigwedge^i$}\, Q \otimes \rho_r =
\bigoplus_{s\in\widehat\Gamma}\, a^{(i)}_{sr}\, \rho_s \qquad
\mbox{with} \quad a^{(i)}_{sr}=\dim_\IC\Hom_\Gamma\big(\rho_s
\,,\,\mbox{$\bigwedge^i$}\, Q \otimes \rho_r \big)
\end{equation}
of the fundamental three-dimensional representation $Q$ of $\Gamma
\subset SL(3,\complex)$, with weights $r_\alpha$, $\alpha=1,2,3$
obeying $r_1+r_2+r_3\equiv0$; explicitly
$a_{rs}^{(1)}=\delta_{r,s+r_1}+\delta_{r,s+r_2}+ \delta_{r,s+r_3}$. To this quiver we associate an ideal of
relations $\langle\sfR\rangle$ defined by $a_{rs}^{(2)}$; for a
Calabi--Yau singularity one has $a_{rs}^{(2)}=a_{sr}^{(1)}$ and
$a_{rs}^{(3)}=\delta_{rs}$. A representation of this quiver is a
$\Gamma$-module which is described by an isotopical decomposition $V =
\bigoplus_{r\in\Gammaw}\, V_r \otimes \rho^\vee_r$ whose factors
correspond to the nodes in the quiver, and a set of linear maps  $B
\in \Hom_{\Gamma} (V , Q \otimes V)$ corresponding to the arrows. By
Schur's lemma these morphisms decompose as
\begin{equation} 
B = \bigoplus_{r\in\Gammaw}\, \big( B_1^{(r)} \,,\, B_2^{(r)} \,,\, B_3^{(r)} \big)
\label{Bdecomp}\end{equation}
where $B_\a^{(r)}\in \Hom_\IC(V_{r}, V_{r+ r_\alpha})$. The ideal of
relations $\langle\sfR\rangle$ imposes conditions on the linear maps
given by the orbifold generalized ADHM equations
\begin{equation}
B_\b^{(r + r_\a)} \ B_\a^{(r)} = B_\a^{(r + r_\b)} \ B_\b^{(r)} \qquad \mbox{for} \quad
\a,\b=1,2,3 \ .
\label{ADHMorb}\end{equation}

The McKay quiver is at the heart of the McKay correspondence. This is
a statement connecting the representation theory data encoded in the
orbifold group $\Gamma$ with the smooth geometry of the natural
crepant resolution $X= \mathrm{Hilb}^{\Gamma} (\complex^3)$. It can be
seen at different levels: as a characterization of the homology and
intersection theory of the resolved geometry in terms of the
representation theory data of $\Gamma$, as a dictionary between the K-theory of $\mathrm{Hilb}^{\Gamma} (\complex^3)$ and the irreducible representations of $\Gamma$, or ultimately (and more deeply) as an equivalence between the derived category of quiver representations of $\sfQ$ and the derived category of coherent sheaves on $\mathrm{Hilb}^{\Gamma} (\complex^3)$.

In the following we will construct  enumerative invariants based on the McKay quiver and use the McKay correspondence to translate our results into geometrical terms.

\section{Stacky gauge theories and their instanton moduli spaces}

We now introduce the concept of a stacky gauge theory and study a moduli space of geometrical objects which are naturally associated with the noncommutative Donaldson--Thomas enumerative problem. A stacky gauge theory is a  sequence of deformations of an ordinary gauge theory on $\complex^3$ whose observables are determined by $\Gamma$-equivariant torsion free $\cO_{\complex^3}$-modules on $\complex^3$, i.e. $\Gamma$-equivariant instantons.

We think of these gauge theories as describing the low-energy dynamics
of D-branes on orbifolds of the form $\complex^3 / \Gamma$ in a
certain ``orbifold phase''. In practice they are realized in the following
way. One starts with ordinary maximally supersymmetric Yang--Mills
theory on $\complex^3$; for the moment we discuss the $U(1)$ gauge theory, but
below we also consider the non-abelian $U(N)$ gauge theory in its Coulomb
branch. This theory is firstly deformed into a noncommutative gauge
theory. Next it is topologically twisted by selecting an appropriate
combination of the supercharges, shifted by inner contraction with the
vector field generating the toric isometries of $\complex^3$, as the
BRST operator. Then the gauge theory localizes onto torus-invariant
noncommutative instantons which were thoroughly studied
in~\cite{Cirafici:2008sn}. These configurations are characterized by
algebraic operator equations on a Fock space ${\mathcal H}$ which have the
ADHM form
\bea \label{ncADHM} \big[Z^{\a}\,,\, Z^{\b}\big] = 0 \ , \qquad
\sum_{\a=1}^3\, \big[Z^{\a}\,,\, Z^{\dagger}_{\a}\, \big] = 3 \qquad \mbox{and} \qquad \big[Z^{\a} \,,\, \Phi\big] = \eps_{\a} \, Z^{\a} \ ,
\eea 
for $\a,\b=1,2,3$.
The operators $Z_{\alpha}$ mix noncommutative coordinates with gauge
field degrees of freedom, $\Phi$ is the Higgs field, and $\eps_\a$ are
equivariant parameters for the natural action of the torus $\torus^3$
on $\IC^3$. These equations can be solved by harmonic oscillator
algebra via the standard creation and annihilation operators
$a_{\alpha}^{\dagger}$, $a_{\alpha}$ on $\mathcal H$ for
$\alpha=1,2,3$. Generic instantons are obtained from these solutions
via partial isometric transformations and correspond to subspaces of the Fock space generated by monomial ideals $I\subset\IC[z^1,z^2,z^3]$ as
$ {\mathcal H}_{I} =  I(a_1^{\dagger}, a_2^{\dagger}, a_3^{\dagger}) | 0,0,0
\rangle$. These ideals are classified by plane partitions $\pi$; the number of boxes of $\pi$ is the instanton charge $k=|\pi|$.

One now considers the orbifold action of $\Gamma$ which is a diagonal subgroup of the torus group $\torus^3\subset SL(3,\IC)$. Then the Fock space of the noncommutative gauge theory is a $\Gamma$-module which decomposes as  
\beq
\cH=\bigoplus_{r\in\widehat\Gamma}\, \cH_{r} \qquad \mbox{with} \quad \cH_{r} = {\rm span}_{\IC}\big\{|n_1,n_2,n_3\rangle~\big|~ n_1\, r_1+n_2\,r_2+n_3\,
r_3\equiv r\big\} \ .
\label{Hisotop}\eeq
As a result the operators $Z_{\alpha}$ decompose as
\beq
Z_\a=\bigoplus_{r\in\Gammaw}\, Z_\a^{(r)} \qquad \mbox{with} \quad
Z_\a^{(r)} \in\Hom_\IC\big(\cH_{r} \,,\, \cH_{r+r_\a}\big)
\label{orbcovcoord}\eeq
and the first of the instanton equations (\ref{ncADHM}) becomes
\beq
Z_\a^{(r+r_\b)}\, Z_\b^{(r)}= Z_\b^{(r+r_\a)}\, Z_\a^{(r)} \ .
\label{orbcomm}\eeq
These holomorphic operator equations are analogous to the matrix equations (\ref{ADHMorb}).
Partial isometries decompose accordingly and the resulting noncommutative instantons are labelled by
$\widehat\Gamma$-coloured plane partitions
$\pi=\bigsqcup_{r\in\widehat\Gamma}\, \pi_r$, where
$(n_1,n_2,n_3)\in\pi_r$ if and only if $n_1\, r_1+n_2\,r_2+n_3\,
r_3\equiv r$.

These gauge theories are associated with a class of moduli spaces
which are obtained in the following way. We interpret the
noncommutative deformation as a desingularization of certain framed moduli spaces of sheaves. These moduli spaces are obtained by applying Beilinson's theorem to a class of torsion free sheaves $\cE$ of rank $N$ and topological charge $\ch_3(\cE)=k$ on the compact toric orbifold $\PP^3 / \Gamma$. This procedure allows us to describe the original sheaf $\cE$ as the single non-vanishing cohomology of a  complex. This complex is characterized by two vector spaces $V$ and $W$ of dimensions $k$ and $N$ which are cohomology groups associated with the original sheaf $\cE$ and which are $\Gamma$-modules, along with a set of tautological bundles. The latter bundles are constructed from the representation theory of $\Gamma$ via the McKay correspondence and characterize the homology of $\mathrm{Hilb}^{\Gamma} (\complex^3)$. In particular the $\Gamma$-module $W$ is associated with the fiber of $\cE$ at infinity. After a rather technical construction one discovers that the relevant moduli spaces can be described in terms of representations of a  quiver. This quiver is the \textit{framed} McKay quiver associated with the orbifold singularity $\complex^3 / \Gamma$. The nodes of this quiver are the vector spaces $V_r$ in the isotopical decomposition of $V$ into irreducible representations of $\Gamma$. The structure of the arrows and relations is precisely that discussed in the previous section.

The only difference now is in the physical interpretation. The dimension $k$ of $V$ represents the instanton number while the dimensions $k_r$ of the individual factors $V_r$ in the decomposition are associated with multi-instantons which transform in the irreducible representation $\rho_r$ (note that this does not imply that each instanton separately is associated with the representation $\rho_r$). The new ingredients are the framing nodes which arise from the isotopical decomposition of the vector space $W=\bigoplus_{r\in\widehat\Gamma}\, W_r\otimes \rho_r^\vee$ into irreducible representations. The framing nodes label boundary conditions at infinity. The gauge fields are required to approach a flat connection at infinity which are classified by the irreducible representations $\rho_r$. At infinity the gauge sheaf is associated with a representation $\rho$ of the orbifold group $\Gamma$ and the
dimensions $\dim_\complex W_r = N_r$ label the multiplicities of
the decomposition of $\rho$ into irreducible representations, with the
constraint
\beq
\sum_{r\in\Gammaw}\, N_r =N \ .
\label{Nsumwr}\eeq
The arrows from the framing nodes correspond to linear maps
$I\in\Hom_\Gamma(W,V)$; by Schur's lemma these morphisms decompose as 
\beq
I = \bigoplus_{r\in\Gammaw}\, I^{(r)}
\label{Idecomp}\eeq
where $I^{(r)}\in\Hom_\complex(W_r,V_r)$.

This construction gives a correspondence between a sheaf $\cE$ (with some technical conditions on its cohomology) preserving certain boundary conditions at infinity and a collection of maps between vector spaces whose algebraic content can be repackaged into a framed McKay quiver. From the complex derived via Beilinson's theorem one can express the Chern character of the original torsion free sheaf $\cE$ in terms of data associated with the representation theory of the orbifold group via the McKay correspondence as
\begin{eqnarray} \label{chE}
\mathrm{ch} (\cE) &=& - \mathrm{ch} \Big( \big( V \otimes \mathcal{R} (-2) \big)^{\Gamma} \Big) +
 \mathrm{ch} \Big(
 \big(\mbox{$V \otimes \bigwedge^2 Q^{\vee}$} \otimes \mathcal{R} (-1) \big)^{\Gamma} \Big)
 \cr & & -\, \mathrm{ch} \Big(
\big(( {V \otimes Q^{\vee} \oplus W}) \otimes \mathcal{R} \big)^{\Gamma}
 \Big) + \mathrm{ch} \Big( \big(  {V} \otimes \mathcal{R} (1) \big)^{\Gamma} \Big) \ .
\end{eqnarray}

The set of tautological bundles 
\begin{equation}
\cR = \bigoplus_{r\in\widehat\Gamma}\, \cR_r \otimes \rho_r
\label{cRdecomp}\end{equation}
when understood geometrically form an integral basis for the K-theory
group $K(X)$ of vector bundles on the resolved space $X= \mathrm{Hilb}^{\Gamma} (\complex^3)$. Furthermore there is a canonical construction which gives two bases $\cV_m$ and $\cR_r$ of $H^4 (X , \zed)$ and $H^2 (X , \zed)$ dual to the bases of exceptional surfaces and curves in the resolution $X$ in terms of linear combinations of Chern classes of the tautological bundles. In the algebraic framework the tautological bundles map to projective objects in the category of quiver representations.

The contribution of each instanton will be assembled into a partition function and weighted by the exponential of the $U(N)$ instanton action
\begin{equation}
S_{\rm inst} = \frac{g_{s}}6 \, \int_{{X}}\, \Tr F_A \wedge F_A \wedge
F_A + \frac12\, \int_{{X}}\, \omega \wedge \Tr F_A \wedge F_A +
\frac1{2g_s}\, \int_{{X}}\, \omega \wedge \omega \wedge \Tr F_A \ ,
\label{Sinstgeneric}\end{equation}
where $g_s$ is the topological string coupling constant.
The exterior products of field strengths $F_A$ can be expressed in
terms of the Chern character $\mathrm{ch} (\cE)$ in (\ref{chE}). Similarly the K\"ahler form $\omega$ on $X$ and its exterior product $\omega \wedge \omega$ can be both expanded in the basis of cohomology determined by the tautological bundles.

The detailed description of the moduli space depends however on the appropriate choice of stability conditions. This boils down to the choice of a stability parameter that enters in the definition of the moduli space. The precise value of that parameter is what will determine which chamber in the moduli space of vacua of the string theory we are working in. To understand properly this issue we now consider a somewhat different perspective via a quantum mechanics associated with the quiver. 

\section{Instanton quantum mechanics and noncommutative Donaldson--Thomas data}

It is customary in instanton computations to use collective coordinates to study the local structure of the moduli space. This corresponds to taking the point of view of the fractional D0-branes which characterize the instantons, in contrast to the point of view of the D6-brane gauge theory we have been considering so far. For this, we will linearize the complex obtained via Beilinson's theorem to construct a local model for the instanton moduli space. This is a rather powerful perspective since to apply toric localization we only need to understand the neighbourhood of each fixed point. 

As we have seen the study of instantons on $\complex^3 / \Gamma$
amounts to an equivariant decomposition of the spaces and maps
involved. One considers the set of bosonic fields (\ref{Bdecomp})
together with (\ref{Idecomp}).
Upon the introduction of the appropriate supermultiplets, the quantum
mechanics is characterized by the generalized ADHM equations
(\ref{ADHMorb}) together with the equations
\begin{eqnarray}
\sum_{\a=1}^{3}\, \Big( B_\a^{(r-r_\a)} \,
B^{(r-r_\a)}_\a \,^{\dagger} - B^{(r)}_\a\,^\dag\,B_\a^{(r)} \Big) + I^{(r)}\,
I^{(r)} \,^{\dagger} = \lambda^{(r)} \ , 
\label{McKayeqs}\end{eqnarray}
where $\lambda^{(r)}>0$; this extra equation is analogous to the second instanton equation of (\ref{ncADHM}). The set of equations (\ref{ADHMorb}) arises as an ideal of relations in the path
algebra of the McKay quiver, while (\ref{McKayeqs}) can be traded for a stability condition. This matrix model is topological and it localizes onto the fixed points of its BRST operator. In the Coulomb phase, these points are classified by
$N$-vectors of plane partitions $\vec \pi = \left( \pi_1 , \dots , \pi_N
\right)$ with $|\vec\pi|=\sum_l\, |\pi_l|=k$ boxes, where each box carries an appropriate $\Gamma$-action. Since the orbifold group $\Gamma$ is a subgroup of the torus group $\torus^3$, the fixed points onto which the matrix quantum mechanics localizes are the same as in the case of the affine space $\complex^3$, the only difference being that one now has to keep track of the $\Gamma$-action.

A local model for the moduli space near a fixed point of the toric action is realized by an equivariant version of the instanton deformation complex
{ \small
\begin{equation} \label{equivdefcomplex}
\xymatrix{
  \Hom_{\Gamma}  (V_{\vec\pi} , V_{\vec\pi})
   \quad\ar[r] &\quad
   {\begin{matrix} \Hom_{\Gamma} (V_{\vec\pi} , V_{\vec\pi} \otimes Q )
   \\ \oplus \\
   \Hom_{\Gamma} (W_{\vec\pi} , V_{\vec\pi}) \\ \oplus  \\ \Hom_{\Gamma} (V_{\vec\pi} ,
   V_{\vec\pi}  \otimes \bigwedge^3 
   Q) \end{matrix}}\quad \ar[r] & \quad
   {\begin{matrix} \Hom_{\Gamma} (V_{\vec\pi} , V_{\vec\pi}  \otimes \bigwedge^2
       Q) \\ \oplus \\ 
       \Hom_{\Gamma} (V_{\vec\pi},W_{\vec\pi} \otimes \bigwedge^3 Q)
   \end{matrix}}
}
\end{equation}
}
from which we can extract the character at the fixed points
\begin{equation} \label{orbcharacter}
\Ch_{\vec\pi}^\Gamma(t_1,t_2,t_3)= \big( W_{\vec\pi}^\vee \otimes V_{\vec\pi} -
{V}_{\vec\pi}^\vee \otimes W_{\vec\pi}+ (1-t_1)\, (1-t_2)\,
(1-t_3) ~ {V}^\vee_{\vec\pi} \otimes V_{\vec\pi} \big)^{\Gamma} \ ,
\end{equation}
where $t_\a=\e^{\ii\eps_\a}$ for $\a=1,2,3$. This yields all the data we need for the construction of noncommutative Donaldson--Thomas invariants.

\section{Enumerative invariants}

To our framed quiver we can associate the representation space
\begin{equation}
\cM^{\Gamma} (\mbf k , \mbf N) =  \Hom_{\Gamma} (V , Q \otimes V) \ 
 \oplus \ \Hom_{\Gamma} (V , \mbox{$\bigwedge^3$} Q \otimes V) \ 
\oplus \ \Hom_{\Gamma} (W , V) \ ,
\end{equation}
where $\mbf k=(k_r)_{r\in\widehat\Gamma}$ and $\mbf N=(N_r)_{r\in\widehat\Gamma}$.
We use the $\Gamma$-equivariant decomposition of the matrix equations
(\ref{ADHMorb}) to define ``moment maps'' $\mu_{\complex}^{\Gamma}$
whose zero locus correspond to the ideal of relations in
the instanton quiver path algebra. These equations define a subvariety
$(\mu_\complex^\Gamma)^{-1}(0) \subset \Hom_{\Gamma} (V , Q \otimes V) \oplus
\Hom_{\Gamma} (V , \bigwedge^3 Q \otimes V)$. This allows us to define
the Donaldson--Thomas quiver moduli space as the quotient stack
\begin{equation}
\frak{M}^\Gamma (\mbf k ,\mbf N) = \Big[ \big((\mu_\complex^{\Gamma})^{-1}(0) \times
\Hom_{\Gamma} (W , V) \big) \, \Big/ \, G_{\mbf k} \Big] \ ,
\end{equation}
where the group
\beq
G_{\mbf k}=\prod_{r\in\widehat\Gamma}\, GL(k_r,\IC)
\eeq
acts by basis change automorphisms of the $\Gamma$-module $V$.
We regard this stack as a moduli space of stable framed
representations in the sense
of~\cite[Section~7.4]{joycesong} when the stability parameter $\mu(\mbf k)=\mu$ defined there takes the value $\mu=0$. 

Noncommutative Donaldson--Thomas invariants are now defined following Behrend \cite{behrend} as the weighted topological Euler characteristics
\begin{equation}
\NDT_{\mu=0} (\mbf k , \mbf N) = \chi \big( \frak{M}^\Gamma (\mbf k ,\mbf N) \, , \, \nu \big) = \sum_{n\in\zed}\, n \
\chi\big(\nu^{-1}(n)\big) \ ,
\label{Behrend}\end{equation}
where $\nu:\frak{M}^\Gamma (\mbf k ,\mbf N) \to\zed$ is a  $G_{\mbf k}$-invariant
constructible function. Our choice of setting the stability parameter $\mu=0$ implies that every object in the category of quiver representations with relations is $0$-semistable. These invariants enumerate
$\Gamma$-equivariant torsion free sheaves on $\complex^3$ via the McKay
correspondence; for ideal sheaves they coincide with the orbifold
Donaldson--Thomas invariants defined in~\cite{young}.

We can construct a partition function for these invariants from the local structure of the instanton moduli space. Neglecting the $\Gamma$-action, the two vector spaces $V$ and $W$ can
be decomposed at a
fixed point $\vec\pi=(\pi_1,\dots,\pi_N)$ of the $U(1)^N \times \torus^3$ action on the
instanton moduli space as~\cite{Cirafici:2008sn}
\begin{eqnarray}
V_{\vec\pi} = \sum_{l=1}^N \,e_l~ \sum_{(n_1,n_2,n_3)\in \pi_l}\,
t_1^{n_1-1} \,t_2^{n_2-1}\,t_3^{n_3-1} \qquad \mbox{and} \qquad W_{\vec\pi} =
\sum_{l=1}^N\,e_l \ ,
\label{decompos}
\end{eqnarray}
where $e_l=\e^{\ii a_l}$ with $a_l$ the Higgs field vacuum expectation values for $l=1,\dots,N$.
Each partition carries an action of $\Gamma$. However this action is offset by the $\Gamma$-action of the factor $e_l$ which corresponds to the choice of a boundary condition on the gauge field at infinity. Recall that the decomposition of $W$ corresponds to imposing boundary
conditions at infinity, which are classified by irreducible
representations of the orbifold group $\Gamma$. In this context each
$U(1)$ factor in the Coulomb phase is associated with a vacuum expectation
value of the Higgs field $a_l$ which corresponds to a certain
irreducible representation of $\Gamma$. Even if the maximal symmetry breaking
pattern $U(N) \rightarrow U(1)^N$ is fixed, one still has to
specify in which superselection sector one is working. This sector is
characterized by choosing which of the eigenvalues $a_l$ are in a particular
irreducible representation of $\Gamma$. The number of eigenvalues of
the Higgs field in the representation $\rho_r^{\vee}$ is precisely
$N_r = \dim_\complex W_r$. Therefore the decomposition of $V_{\vec\pi}$ can be also written as
\begin{equation}
V_{\vec\pi} =\bigoplus_{l=1}^N ~ \bigoplus_{r\in\Gammaw} \, \big( E_l \otimes \rho_{b(l)}^{\vee} \big) \otimes \left( P_{l,r} \otimes \rho_r^{\vee} \right) =\bigoplus_{l=1}^N~ \bigoplus_{r\in\Gammaw} \, \big( E_l \otimes P_{l,r} \big) \otimes \rho_{r+b(l)}^{\vee}
\label{VpiGammadecomp}\end{equation}
where $E_l$ is the $\Gamma$-module generated by $e_l$, and we have introduced
the boundary function $b(l)$ which to each sector $l$ corresponding to
a module $E_l$ associates the weight of the corresponding
representation of $\Gamma$; if the vacuum expectation value $e_l$
transforms in the irreducible representation $\rho_s$, then
$b(l)=s$. Here $P_{l,r}$ are vector spaces which appear in the
$\Gamma$-module decomposition of the sum $\sum_{(n_1,n_2,n_3)\in \pi_l}\, t_1^{n_1-1} \,t_2^{n_2-1}\,t_3^{n_3-1} $. From this formula one can derive a relation between the instanton numbers and the number of boxes in a partition associated with a given irreducible representation; it is given by
\begin{equation} \label{NAinst}
k_r = \sum_{l=1}^N \, |\pi_{l,r-b(l)}| \ .
\end{equation}

The contribution of an instanton to the gauge theory fluctuation
determinant can be now derived from the local character
(\ref{orbcharacter}) of the moduli space near a fixed point; it is given by $(-1)^{{\mathcal K} (\vec\pi;\mbf N)} $, with
\begin{eqnarray}
\cK(\vec\pi;\mbf N) &=& \sum_{l=1}^N ~ \sum_{r\in\Gammaw}\, |\pi_{l,r}| \ N_{r+b(l)}  
\nonumber \\ &&
- \sum_{l,l'=1}^N \ \sum_{r\in\Gammaw}\, |\pi_{l,r}|\, \Big( |\pi_{l',r+b(l)-b(l'\,)-r_1-r_2}| - |\pi_{l',r+b(l)-b(l'\,)-r_1}| \nonumber \\ &&  -\, |\pi_{l',r+b(l)-b(l'\,)-r_2}| + |\pi_{l',r+b(l)-b(l'\,)}| \Big) \ .
\label{instmeasure}\end{eqnarray} 
The fixed point values of the instanton action (\ref{Sinstgeneric}) in these variables can be written as
{ \small
\begin{eqnarray}
S_{\rm inst}(\vec\pi;\mbf N)  \nonumber \hspace{-1.6cm} && \\ &=& -\frac1{2g_s} \ 
\sum_{m,r,s\in\Gammaw}\, \varsigma_m \, \Big( N_s\,\delta_{rs} -
\big(  a^{(2)}_{rs} - a^{(1)}_{rs} \big)\, \sum_{l=1}^N\,
|\pi_{l,s-b(l)}| \Big)\, \int_{X}\, c_2 (\cV_m) \wedge c_1 ({\cR_r})
 \nonumber \\ && 
+\, \sum_{n,r,s\in\Gammaw}\, \varphi_n\, \bigg( \Big( N_s\, \delta_{rs} - \big( a^{(2)}_{rs} - a^{(1)}_{rs} \big) \, \sum_{l=1}^N \, |\pi_{l,s-b(l)}| \Big)\, \int_{X}\, c_1 (\cR_n) \wedge \mathrm{ch}_2 ({\cR_r}) \cr && \qquad \qquad +\,\big(  a_{rs}^{(2)}  - 3 \delta_{rs} \big) \,  \sum_{l=1}^N\, |\pi_{l,s-b(l)}| \, \int_{X}\, c_1 (\cR_n) \wedge c_1 \big(\cO_{X}(1)\big) \wedge c_1(\cR_r)
  \bigg) \nonumber \\ &&
 -\,g_s\, \sum_{r,s\in\Gammaw}\, \bigg(\Big( N_s\,\delta_{rs} - \big( a^{(2)}_{rs} - a^{(1)}_{rs} \big) \, \sum_{l=1}^N\, |\pi_{l,s-b(l)}| \Big)\, \int_{X}\, \mathrm{ch}_3 ({\cR_r}) \cr && \qquad \qquad
  +\,\big( a_{rs}^{(2)} - 3 \delta_{rs} \big) \, \sum_{l=1}^N\, |\pi_{l,s-b(l)}| \, \int_{X}\, c_1 \big(\cO_{X}(1)\big) \wedge \ch_2 (\cR_r)
  \cr & & \qquad \qquad +\, \big(  a^{(2)}_{rs} - 3 \delta_{rs} \big)
  \, \sum_{l=1}^N \, |\pi_{l,s-b(l)}| \, \int_{X}\,  c_1
  (\cR_r) \wedge \ch_2 \big(\cO_{X} (1)\big) \bigg) \cr &&
  +\, \frac{g_s}{|\Gamma|}\, \sum_{s\in\Gammaw} \ \sum_{l=1}^N \, |\pi_{l,s-b(l)}|  \ ,
\label{ch3Naction} \end{eqnarray} }
where $\varphi_n$ (resp. $\varsigma_m$) are chemical potentials for the D2-branes (resp. D4-branes) determined by the expansion of $\omega$ (resp. $\omega\wedge\omega$) into the basis of tautological bundles $\cR_n$ (resp. $\cV_m$).
Note that the choice of boundary condition enters not only explicitly in the dimensions $N_r$, but also implicitly in the plane partitions.

Finally, the partition function for noncommutative Donaldson--Thomas invariants of type $\mbf N$ is in full generality given by
\begin{equation}
\cZ_{\complex^3 / \Gamma}\left(\mbf N\right) = \sum_{\vec \pi} \, (-1)^{\cK(\vec\pi;\mbf N)}~ \e^{-S_{\rm inst} (\vec \pi ; \mbf N)} \ .
\end{equation}
The instanton action is naturally rephrased in terms of intersection
indices on the homology of the crepant resolution $X= \mathrm{Hilb}^{\Gamma} (\complex^3)$, via the McKay correspondence. However it is computed via the instanton numbers that characterize the noncommutative Donaldson--Thomas invariants, which are the relevant variables in the noncommutative crepant resolution chamber. See~\cite{Cirafici:2010bd} for various explicit examples and applications of this formalism.

\section{BPS invariants}

The noncommutative invariants described in the previous section are related to the quiver generalized Donaldson--Thomas invariants $\DT_\mu(\mbf k) \in \mathbb{Q}$ defined by Joyce and
Song in~\cite{joycesong} through
{ \small
\bea
\NDT_\mu(\mbf k, \mbf N) &=& \sum_{m=1}^\infty~
\sum_{\stackrel{\scriptstyle \mbf k_1,\dots, \mbf k_m\neq\mbf
    0}{\scriptstyle \mbf k_1+\cdots + \mbf k_m=\mbf k \ , \ \mu(\mbf
    k_i)=\mu(\mbf k)}} \, \frac{(-1)^m}{m!} \label{NCDTgenrel} \\ && \hspace{-2cm}
    \times\, \prod_{i=1}^m\, \Big((-1)^{\mbf k_i\cdot \mbf
      N-\langle\mbf k_1+\cdots+\mbf k_{i-1},\mbf k_i\rangle}\, \big(\mbf
    k_i\cdot \mbf N-\langle\mbf k_1+\cdots+\mbf k_{i-1},\mbf k_i \rangle
    \big)\, \DT_\mu(\mbf k_i)\Big)
\nonumber \eea }
where the skew-symmetric bilinear form
$\langle -,-\rangle:\IN_0^{|\Gamma|}\times\IN_0^{|\Gamma|} \to\IZ$ given by
\beq
\langle \mbf k,\mbf k'\, \rangle =
\sum_{r,s\in\widehat\Gamma}\,\big(a_{sr}^{(1)}-
a_{rs}^{(1)}\big)\, k_r\, k_s'
\label{chibar}\eeq
is the antisymmetrization of the Euler--Ringel form of the quiver $\sfQ$.
In the case of semi-small crepant
resolutions, the forms (\ref{chibar}) vanish and this equation
yields a useful relationship between the corresponding partition
functions
\begin{equation} \label{quivergeneralized}
1 + \sum_{\mbf k \, : \, \mu(\mbf k)=\mu}\, \NDT_\mu(\mbf k,\mbf N) \,
\mbf p^{\mbf
  k} = \exp\Big( - \sum_{\mbf k \, : \, \mu(\mbf k)=\mu}\, (-1)^{\mbf k
  \cdot \mbf N}\, \left( \mbf k \cdot \mbf N \right) \,
\DT_\mu(\mbf k) \, \mbf p^{\mbf k} \Big)
\end{equation}
where $\mbf p^{\mbf k}:= \prod_{r\in\Gammaw}\, p_r^{k_r}$. This shows that the structure captured by the noncommutative invariants is encoded, perhaps more fundamentally, in the quiver generalized Donaldson--Thomas invariants. Furthermore from these invariants one can define the quiver BPS invariants $\BPS_\mu(\mbf k) \in \mathbb{Q}$ as
\begin{equation} \label{mob}
\BPS_\mu(\mbf k) = \sum_{ m \ge
    1 \, : \, m | \mbf k }\, \frac{\text{M\"o}(m)}{m^2} \
\DT_\mu(\mbf k/m) \ ,
\end{equation}
which are conjectured to count BPS states; here $\text{M\"o}:\mathbb{N}\to\mathbb{Q}$ is the M\"obius function. Our formalism provides a solid ground to assess this conjecture since all these invariants are in principle computable from the noncommutative Donaldson--Thomas partition functions. These computations are reduced to the combinatorial problem of counting plane partitions while keeping track of the $\Gamma$-action. Again see~\cite{Cirafici:2010bd} for some explicit examples. Physically (\ref{mob}) was interpreted in \cite{Manschot:2010qz} as an effective degeneracy which allows to treat the constituents of multi--centered bound states as Maxwell--Boltzmann particles.

Let us briefly discuss the D-brane interpretation of this picture. The
noncommutative invariants depend on a pair of vectors of integers $\left( \mbf k , \mbf N \right)$. The vector $\mbf k$ labels the instanton numbers and contains the information about which instanton configuration is associated with a given representation of $\Gamma$, though not directly but in the way we have explained in the previous section. On the other hand the vector $\mbf N$ labels boundary conditions for the gauge sheaf. If for simplicity we consider the $U(1)$ theory then the framing of the McKay quiver only adds one extra node. This node, corresponding to the D6-brane, can be connected to any of the nodes of the original quiver. Since each node corresponds to an irreducible representation, this choice reflects how the information about the boundary condition is encoded in the quiver. In the language of \cite{Ooguri:2008yb} the position of the extra nodes determines how cyclic modules are based and therefore the particular enumerative problem. In our picture the reason for this is clear: the choice of the reference node corresponds to a superselection sector in the space of states of the worldvolume gauge theory. In particular, the numerical value of the noncommutative Donaldson--Thomas invariants will be \textit{different} in each sector. However thanks to the formulas (\ref{NCDTgenrel}) and (\ref{quivergeneralized}) all these invariants are \textit{equivalent}, i.e. they can all be expressed in terms of the same set of invariants $\DT_{\mu=0}(\mbf k)$ which are independent of the boundary conditions; the dependence on the vector $\mbf N$ is completely encoded in the prefactors. This is in perfect agreement with our physical expectations of the noncommutative invariants~$\NDT_{\mu=0}(\mbf k,\mbf N) $. 

\section{Combinatorics of orbifold partitions and refined invariants\label{sec:refined}}

The counting of BPS states at orbifold singularities has an underlying combinatorial problem. This is essentially the classical melting crystal problem of \cite{Okounkov:2003sp} with two modifications. Firstly there is a colouring of each partition which is uniquely specified by the orbifold action, secondly each configuration has the sign weight (\ref{instmeasure}). 
There are two ingredients that enter into the colouring of a
partition: the orbifold action directly on the instanton vector spaces
$P_{l,r}$ and the overall shift determined by the boundary conditions,
see (\ref{VpiGammadecomp}). This produces in general a quite intricate
combinatorial problem depending on the boundary conditions. 
Despite the sign factor, which would appear to cancel BPS state contributions corresponding to coloured partitions of different shape but equal colour charges, this combinatorial prescription is still
related to a three-dimensional melting crystal model but with an additional weighting parameter. Note that, exactly as it happens in \cite{Cirafici:2010bd}, the stacky gauge theory formalism predict a form for the BPS states partition function which is different from those considered in the literature, for example in \cite{young}. In particular, while generically in the mathematics literature one is interested in a combinatorial partition function and is therefore natural to introduce parameters associated with the colouring of the partition, here we don't have any freedom and the combinatorial configurations have to be weighted by the instanton action. Physically this is a consequence of the fact that the correct BPS generating function is a sum over sectors of fixed BPS charge. We discuss the relation between the two kinds of partition functions.

We use our formalism to define ``refined'' invariants by
adapting the combinatorial arguments
of~\cite{Iqbal:2007ii,Dimofte:2009bv}. For fixed boundary conditions, our partition functions have the schematic forms~\cite{Cirafici:2010bd}
\begin{equation}
\mathcal{Z}_{\IC^3/\Gamma}(\mbf N) = \sum_{\vec\pi}\,
(-1)^{\mathcal{K}(\vec\pi;\mbf N)}\, q^{\ch_3 (\cE_{\vec\pi_b})}\ \prod_{n=1}^{b_2}\,
Q_n^{\ch_2 (\cE_{\vec\pi_b})_{n}}\ \prod_{m=1}^{b_4}\, U_m^{c_1(\cE_{\vec\pi_b})_m} \ ,
\label{Zschem}\end{equation}
where $\vec\pi_b:=\big(\pi_{1,r-b(1)},\dots,\pi_{N,r-b(N)} \big)_{r\in\Gammaw}$.
The product over the variables $Q_n=\e^{-\varphi_n}$ (resp. $U_m=\e^{-\varsigma_m/2g_s}$) corresponds to the number of
generators $b_2$ (resp. $b_4$) of the resolved homology $H_2 (X,\IZ)$
(resp. $H_4(X,\IZ)$); when $X={\rm Hilb}^\Gamma(\complex^3)$ is a
semi-small crepant resolution one has $c_1(\cE_{\vec\pi_b})=0$. The
parameter $q=\e^{-g_s}$ is weighted by the third Chern characteristic class of the
fixed-point sheaf $\cE_{\vec\pi_b}$ expressed in terms of coloured partitions
in (\ref{ch3Naction}); in the case
$X=\complex^3$ and $N=1$ this would correspond to the total number of boxes
$|\pi|$ in the plane partition~$\pi$.

This counting can be equivalently recast in terms of two-dimensional
partitions obtained by slicing a three-dimensional partition
$\pi$. The two-dimensional partitions interlace each other. If
we think of drawing a plane partition $\pi$ in the space $(x,y,z)$, then
this slicing can be done in such a way that the two-dimensional
partitions $\pi(a)$ live on planes $x-y=a$ with $a \in \zed$
and $\sum_{a\in\zed}\, |\pi(a)|=|\pi|$~\cite{Iqbal:2007ii}. This definition is independent of the colouring of the partition. To ``refine'' our counting we can weigh the slices with $a\geq0$ and $a < 0$ differently, but independently of the colouring of the partition, with parameters $q_1$ and $q_2$ respectively. The resulting partition function has the schematic form
\begin{eqnarray}
\mathcal{Z}_{\IC^3/\Gamma}^{\rm ref}(\mbf N) &=& \sum_{\vec\pi}\,
(-1)^{\mathcal{K}(\vec\pi;\mbf N)}\, \Big(\, \prod_{a=1}^\infty\, q_1^{\ch_3
  (\cE_{\vec\pi_b(a-1)})}\, q_2^{\ch_3 (\cE_{\vec\pi_b(-a)})} \, \Big) \  \nonumber \\ && \hspace{2cm} \times
\prod_{n=1}^{b_2}\, Q_n^{\ch_2 (\cE_{\vec\pi_b})_{n}}\ \prod_{m=1}^{b_4}\,
U_m^{c_1 (\cE_{\vec\pi_b})_{m}} \ .
\label{Zrefschem}\end{eqnarray} 

We write the counting weights as
\beq
q_1=q\, \lambda \qquad \mbox{and} \qquad q_2=q\, \lambda^{-1} \ .
\label{qlambdaweights}\eeq
The ``classical'' limit is $\lambda=1$,
$q_1=q_2=q$. In~\cite{Iqbal:2007ii} the extra parameter $\lambda$ makes
the graviphoton background non-selfdual, and so accounts for the
second $SU(2)$ factor of the spatial rotation group of $\real^4$,
i.e. for the spin content of the D6--D4--D2--D0 bound states on $X$;
in~\cite{Dimofte:2009bv} it is identified with the square root of the
Lefschetz motive of the affine line $\complex$ in motivic
Donaldson--Thomas theory~\cite{Kontsevich:2009xt}. As shown
in~\cite{Cirafici:2010bd}, there exists a simple change of variables
$(q,Q_n,U_m)\mapsto (p_r)_{r\in\Gammaw}$ from the large radius
parameters in (\ref{Zschem}) to orbifold parameters
$p_r$ which weigh plane partitions $\pi_r$ of colour $r$ with
\beq
\prod_{r\in\Gammaw}\, p_r=q \ .
\eeq
Then the refined
partition function (\ref{Zrefschem}) takes the form
\beq
\mathcal{Z}_{\IC^3/\Gamma}^{\rm ref}(\mbf N) = \sum_{\vec\pi}\,
(-1)^{\mathcal{K}(\vec\pi;\mbf N)}\, \lambda^{2s_{\vec\pi}} \ \prod_{r\in\Gammaw}\,
p_r^{\sum_{l=1}^N\, |\pi_{l,r-b(l)}|} \ ,
\label{Zreforb}\eeq
where the spin content $s_{\vec\pi}$ of the $\Gamma$-equivariant instantons on
$\IC^3$ is captured by the sum over intersection indices
{ \small
\bea
s_{\vec\pi} &=& \frac12 \, \sum_{a=1}^\infty \ \sum_{r,s\in\Gammaw} \
\sum_{l=1}^N \,
\Big(\, \big|\pi_{l,r-b(l)}(a-1)\big|-\big|\pi_{l,r-b(l)}(-a) \big|\, \Big) \, 
\\ \nonumber 
&& \times 
\bigg(\, \frac1{|\Gamma|} + \big(a_{sr}^{(1)}-a_{rs}^{(1)} \big)
\ \int_{X}\,
\mathrm{ch}_3 ({\cR_s})  \nonumber \\ && \nonumber \hspace{-0.2cm}
\qquad 
+\,\big(
a_{sr}^{(1)} - 3 \delta_{rs} \big) \ \int_X\, \Big( c_1
  (\cR_s) \wedge \ch_2 \big(\cO_{X} (1)\big) - c_1
  \big(\cO_{X}(1)\big) \wedge \ch_2 (\cR_s) \Big) \bigg) \ ,
\eea  }which in several cases can be worked out explicitly by using the
calculations of~\cite{Cirafici:2010bd}. The refined
noncommutative invariants
generated by this partition function will be denoted $\NDT^{\rm ref}_{\mu=0}
(\mbf k , \mbf N;\lambda)$; the ordinary (unrefined) invariants are
recovered in the classical limit $\lambda=1$ as $\NDT_{\mu=0}
(\mbf k , \mbf N)= \NDT^{\rm ref}_{\mu=0}
(\mbf k , \mbf N;1)$.

In the affine case $X=\IC^3$, this partition function is a higher-rank
version of the refined MacMahon function~\cite{Iqbal:2007ii} in the
Coulomb branch given by
\bea
\mathcal{Z}_{\IC^3}^{\rm ref}(\mbf N) &=& \sum_{\vec\pi}\,
(-1)^{N\,|\vec\pi|}\ \lambda^{\sum_{a\in\IN}\,
  (|\vec\pi(a-1)|-|\vec\pi(-a)|)} \ q^{|\vec\pi|} \nonumber \\[4pt]
&=& \prod_{n=1}^\infty \ \prod_{k=1}^{n} \, \big(1-(-1)^{N\, n} \ \lambda^{2k-n} \ q^n \big)^{-N} \ ,
\eea
which refines the partition functions of~\cite{Cirafici:2008sn}. 
It would be interesting to further examine the physics behind these
combinatorial definitions, for example their relations to the
introduction of probe branes in our orbifold construction, and how
this construction is related to the quiver BPS invariants, perhaps
along the lines of \cite{Ooguri:2008yb}.

\section{Virtual instanton quivers and charges}

In our formalism there is a natural basis of quantum BPS states given
by the basis of fractional D0-branes, bound to either a single
D6-brane or to multiple D6-branes. This basis is identified with the simple representations of the quiver $\sfD_r$ and each basis element naturally corresponds to a node of the quiver. To this basis we must add the generators of D6-brane charge. We will describe this lattice of charges via the McKay correspondence. The correspondence suggests to endow this lattice with the intersection pairing in $K^c(X)$. With this procedure we associate to any instanton quiver a ``virtual" quiver. Physically this virtual quiver can be obtained from the instanton quiver matrix quantum mechanics by introducing gauge invariant mass terms for all the oriented 2--cycles and then decoupling the relevant fields by taking their masses to be large.

Our quivers have the generic form of a collection of nodes associated
with fractional branes that may be connected with auxiliary framing
nodes, which represent the D6-branes and contain all the relevant
information about the boundary conditions. To this structure we can
naturally associate a lattice $\Lambda$, which we call the
\textit{lattice of fractional brane charges}. We take the fractional
D0-branes as generators for this lattice; this would appear to identify
$\Lambda$ as the lattice of K-theory charges $K(X)$, but below we shall argue
that it is more appropriate to use the dual K-theory group $K^c(X)$ of
\emph{complexes} of vector bundles on the resolution $X$ which are
exact outside the exceptional locus. Therefore our charges are
naturally labelled by the irreducible representations
$r\in\widehat\Gamma$ of the orbifold group associated with the
original nodes of the quiver and we will denote them as
$\gamma_{r}$. We then add an extra generator for each framing node,
corresponding to the D6-brane charge and denoted $\bullet$, and call the corresponding generators $\gamma_{\infty}$. For simplicity we will usually consider configurations with total D6-brane charge equal to one, i.e. a single D6-brane labelling trivial boundary conditions at infinity. We will collectively denote the set of generators with $\gamma_{I}$ where the index $I$ runs over the irreducible representations $r\in\widehat\Gamma$ and the framing nodes~$\bullet$.

We would like to endow the lattice $\Lambda$ thus defined with a
skew-symmetric bilinear intersection pairing
$(-,-):\Lambda\times\Lambda\to \zed$. A natural choice would be a
pairing dictated by the arrow structure of the quiver, in analogy with
the construction of~\cite{Cecotti:2010fi}. We will see below that a proper pairing actually takes into account the relations of the McKay quiver as well.

Consider for example the singularity $\complex^3 / \zed_3$. The McKay quiver is determined by the tensor product decomposition (\ref{tensordecomp}), which in this case is given by~\cite{Cirafici:2010bd}
\begin{equation}
 a_{rs}^{(1)}= \left( \begin{matrix} 0 & 0 & 3 \\ 3 & 0 & 0 \\  0 & 3 & 0 \end{matrix} \right)
 \qquad \mbox{and} \qquad 
 a_{rs}^{(2)} = \left( \begin{matrix} 0 & 3 & 0 \\ 0 & 0 & 3 \\  3 & 0 & 0 \end{matrix} \right)
\label{C3Z3matrices}\end{equation}
with $a_{rs}^{(2)}= a_{sr}^{(1)}$. Therefore the instanton quiver, in the case of trivial boundary condition, has the form
\begin{equation}
\vspace{4pt}
\begin{xy}
\xymatrix@C=20mm{ & \ W_0 \ \bullet \ar[d] & \\
& \ V_0 \ \circ \ \ar@/^/[ddl] \ar@/_0.5pc/[ddl] \ar@//[ddl]  & \\
& & \\
V_1 \ \circ \ \ar@//[rr] \ar@/^/[rr]  \ar@/_/[rr]   & &  \ \circ \ V_2  \ar@/^/[uul] \ar@/_0.5pc/[uul] \ar@//[uul] 
}
\end{xy}
\vspace{4pt}
\label{quiverC3Z3}\end{equation}
with representation provided by the vector spaces $V_r$, $r=0,1,2$ that enter in the decomposition of $V$ into \textit{dual} irreducible representations of $\Gamma$. 

The first matrix in (\ref{C3Z3matrices}) contains the information about the arrow structure; the matrix element $a^{(1)}_{rs}$ is the number of arrows from node $r$ to node $s$. Then we get a skew-symmetric pairing by setting
\begin{eqnarray}
 \left( \gamma_{r} , \gamma_{s} \right) = a^{(1)}_{rs} \qquad \mbox{and} \qquad
  \left( \gamma_{s} , \gamma_{r} \right) = - a^{(2)}_{sr} \ .
\end{eqnarray}
Thus the matrix of charge pairings is the antisymmetrization of the matrix $a_{rs}^{(1)}$. The extension of this definition to include the framing vertices is immediate. For example for the $U(1)$ gauge theory with trivial boundary condition one sets
\begin{equation}
\left( \gamma_{\infty} , \gamma_{0} \right) = - \left( \gamma_{0} , \gamma_{\infty} \right) = 1
\end{equation}
and $(\gamma_\infty,\gamma_r)=0=(\gamma_r,\gamma_\infty)$ for $r\neq0$. In this case the skew-symmetric pairing between charges is given by the matrix
\begin{equation} \label{pairingC3Z3}
\left( \gamma_{I} , \gamma_{J} \right) = \left( \begin{matrix}0 & 1 & 0 & 0 \\ -1 &  0 & 3 & -3 \\ 0 & -3 & 0 & 3 \\ 0 &  3 & -3 & 0 \end{matrix} \right) \ .
\end{equation}
This pairing is non-degenerate because the matrix (\ref{pairingC3Z3}) has determinant equal to $9$.

However in the general case we have to modify this pairing, because a
generic McKay quiver has oriented 2-cycles (i.e. closed paths of the form $
\begin{xy}
\xymatrix@C=10mm{ \circ \ar@/^/[r]  &  \ar@/^/[l]   \circ
}
\end{xy}
$) which translate into a partial symmetry of the matrix
$a_{rs}^{(1)}$. In order to have a totally skew-symmetric pairing
these 2-cycles have to be removed. We will now see that the McKay quivers have a natural skew-symmetric pairing associated via the McKay correspondence with the intersection theory of the resolved singularity.
 
It is always possible to associate to the instanton quiver a
``virtual'' quiver which is 2-acyclic, because one has naturally
associated to it a perfect intersection pairing on $K^c (X)$ given on
a basis $\cS_r$ dual to $\cR_r$ by the index~\cite{Cirafici:2010bd}
\begin{equation}
\left( \cS^{\vee}_r \,,\, \cS_s  \right) := \int_X\, \ch\big(\cS^{\vee}_r
\otimes \cS_s \big) \wedge{\rm Todd} (X) =a^{(2)}_{rs} - a^{(1)}_{rs} \ .
\end{equation}
The rationale behind this choice of pairing is that we can identify the
complexes of vector bundles $\cS_r$ on $X$ with states of D-branes which
naturally correspond to fractional 0-branes (equivalently
$\Gamma$-equivariant instantons on $\IC^3$), and after
the framing with BPS bound states of D-branes. It is therefore more physical and expected to reproduce in the large radius limit the geometrical pairing between electrically and magnetically charged D-branes. We shall denote this pairing in the fractional brane lattice $\Lambda$ as
\begin{equation} \label{Vpairing}
\langle \gamma_{r} , \gamma_{s} \rangle = a^{(2)}_{rs} - a^{(1)}_{rs} \ ,
\end{equation}
which can be extended in the obvious way to include also framing nodes representing D6-branes. This pairing coincides with
(\ref{chibar}) for the basis of fractional D0-branes given by
$\gamma_r=(\delta_{rs})_{s\in\Gammaw}$ for all $r\in\Gammaw$. We
define the \textit{virtual} quiver associated to a Calabi--Yau
orbifold singularity $\complex^3 / \Gamma$ as the quiver whose nodes
are the same as those of the McKay quiver but whose adjacency matrix
is given by the pairing~(\ref{Vpairing}).

The resulting virtual quiver
is always 2-acyclic: for a Calabi--Yau singularity one has
$a^{(2)}_{rs} = a^{(1)}_{sr}$, and therefore the new pairing $\langle
- , - \rangle$ is the antisymmetrization of the matrix $a_{rs}^{(1)}$,
which automatically removes any oriented 2-cycles in the quiver which
arise from the symmetric part of $a_{rs}^{(1)}$. We will denote by
$\widehat{\Lambda}$ the lattice of charges $\Lambda$ with this new
pairing between fractional branes (extended to include framing
nodes). There is no guarantee that the new pairing is non-degenerate. We can cure the degeneracy problem by embedding the lattice $\widehat{\Lambda}$ into a larger symplectic lattice as explained in~\cite[Section 2.6]{Kontsevich:2009xt}. A natural choice is $\widehat{\Lambda} \oplus \widehat{\Lambda}\,^{\vee}$, where the dual lattice is defined as $\widehat{\Lambda}\,^\vee = \Hom_\zed(\,\widehat{\Lambda} , \zed)$, endowed with the pairing
\begin{equation}
\big\langle (\gamma_{I} , \nu_{J}) \, , \, (\gamma_{{K}} , \nu_{{L}}) \big\rangle = \langle \gamma_{I} , \gamma_{J} \rangle +  \nu_{J} (\gamma_{{K}} ) - \nu_{{L}}  (\gamma_{I} ) \ .
\end{equation}
In the following we will implicitly assume that the charge lattice and the pairing are, if necessary, enlarged in this way.

For example, in the case of $\complex^3 / \zed_3$ the pairing is still given by (\ref{pairingC3Z3}). Things are however slightly different for other singularities such as $\complex^3 / \zed_6$; neglecting the framing nodes for a moment, in this case the adjacency matrix and the matrix of relations of the McKay quiver are~\cite{Cirafici:2010bd}
\begin{equation}
a^{(1)}_{rs} = \left( \begin{matrix} 
0 & 0 & 0 & 1 & 1 & 1 \\
1 & 0 & 0 & 0 & 1 & 1 \\
1 & 1 & 0 & 0 & 0 & 1 \\
1 & 1 & 1 & 0 & 0 & 0 \\
0 & 1 & 1 & 1 & 0 & 0 \\
0 & 0 & 1 & 1 & 1 & 0 \\
\end{matrix} \right)
\qquad \mbox{and} \qquad
a^{(2)}_{rs} = \left( \begin{matrix} 
0 & 1 & 1 & 1 & 0 & 0 \\
0 & 0 & 1 & 1 & 1 & 0 \\
0 & 0 & 0 & 1 & 1 & 1 \\
1 & 0 & 0 & 0 & 1 & 1 \\
1 & 1 & 0 & 0 & 0 & 1 \\
1 & 1 & 1 & 0 & 0 & 0 \\
\end{matrix} \right) \ .
\end{equation}
On the other hand the pairing (\ref{Vpairing}) between fractional branes is
\begin{equation} \label{intC3Z6}
\langle \gamma_{{r}} , \gamma_{s} \rangle = \left( \begin{matrix} 
0 & 1 & 1 & 0 & -1 & -1 \\
-1 & 0 & 1 & 1 & 0 & -1 \\
-1 & -1 & 0 & 1 & 1 & 0 \\
0 & -1 & -1 & 0 & 1 & 1 \\
1 & 0 & -1 & -1 & 0 & 1 \\
1 & 1 & 0 & -1 & -1 & 0 \\
\end{matrix} \right) \ ,
\end{equation}
which corresponds to the virtual instanton quiver
\begin{equation}
\vspace{4pt}
\begin{xy}
\xymatrix@C=20mm{
& \ V_0 \ \circ \ \ar@/^/[dr] \ar@/^0.5pc/[ddr] & \\
V_5 \ \circ \ \ar@/^/[ur]  \ar@/^/[rr] & & \ V_1 \ \circ \ar@/^/[d] \ar@/^/[ddl]  \\
V_4 \ \circ \ \ar@/^/[u] \ar@/^/[uur]    & &  \ V_2 \ \circ \ar@/^/[dl] \ar@/^/[ll] \\
& \ V_3 \ \circ \ \ar@/^/[ul] \ar@/^/[uul] & 
}
\end{xy}
\vspace{4pt}
\label{C3Z6virtual}\end{equation}
to which one can add the framing nodes.

The lattice $\widehat{\Lambda}$ just introduced is somewhat different
from the physical lattices associated with D-branes wrapping cycles in
large volume compactifications. There is no immediate splitting into
electric and magnetic charges in the noncommutative crepant
resolution. The usual electric-magnetic splitting in large radius
variables is described by an intersection pairing that always takes
the schematic form
\begin{equation}
\left\langle\!\left\langle (e,m) \,,\, (e',m'\,) \right\rangle\!\right\rangle = e\, m' - e'\, m \ ,
\end{equation}
which reflects the physical expectations of electrically and magnetically charged particles; for example, two electrically charged particles are always mutually local, and so on. These properties are somewhat hidden in our new pairing dictated by the McKay correspondence, although it passes the obvious physical requirement that a state is local with itself (the pairing then vanishes since it is skew-symmetric). It would be very interesting to study the pattern of local and non-local charges associated with our pairing; unfortunately this question is somewhat tricky as one has also to specify if a moduli space is empty or not for each given orbifold singularity. This appears at first sight to be a formidable problem and we hope to address it in a future work.

For now let us note that there is a natural electric-magnetic splitting. While the charge of the D6-brane is fixed to one, D0-branes are easily interpreted as regular instantons, i.e. configurations of instantons which are symmetric under the orbifold group action and which are therefore free to move off the orbifold singularity. On the other hand D2-branes and D4-branes correspond respectively to elements of $H_2 (X,\zed)$ and $H_4 (X,\zed)$ to which we can canonically associate a dual basis in cohomology given by the Chern classes of the tautological bundles. Therefore we can canonically associate to these states the dual (now in the K-theory sense) elements $\cS_r$. Altogether this constitutes a natural splitting between electric and magnetic charges which naively corresponds to the usual electric-magnetic splitting in the cohomology lattice $H^{\sharp} (X , \zed)$.

\section{Motivic invariants from instanton quivers\label{sec:motivic}}

Our instanton quivers can be also used to set up the problem of computing
motivic Donaldson--Thomas invariants for abelian quotient singularities, which are closely related to the
refined invariants described in Section~\ref{sec:refined} and will
naturally set up our discussion of wall-crossing in subsequent sections. There are
several results in the theory of motivic invariants on local
threefolds, see
e.g.~\cite{bbs,mozgovoy1,nagao,morrison,MMNSz,mozgovoy2}, following
the seminal works~\cite{Kontsevich:2009xt,Kontsevich:2010px}. We adapt these results to our formalism and extend them to the Coulomb branch invariants $\big[\NDT_{\mu=0} (\mbf k , \mbf N)\big] $ which depend on the choice of a boundary condition. The
results of this section can be straightforwardly extended also to the
virtual quivers, although for simplicity we will only consider the
stable instanton quivers. We will begin with some generalities. In what follows we will be rather
sketchy in the formal details (for which we refer the reader to the
review~\cite{bridgeland}) and focus more on computational
aspects. 

The idea behind enumerative problems is to count invariants
associated with moduli spaces of BPS states. There is however
a meaningful way to associate an enumerative problem to the moduli
spaces themselves. For this, one defines the abelian K-theory group of
varieties which is generated by isomorphism classes of complex varieties
$\frM$ 
modulo the scissor relations $
[\frM] = [\frZ] + [\frM \setminus \frZ]$
whenever $\frZ$ is a subvariety of $\frM$. The group structure is
given by $[\frM] + [\frN] = [\frM \sqcup \frN]$. A commutative ring structure comes
from setting $[\frM]\, [\frN]= [\frM \times \frN]$; the class of the
point is the unit $1 = [\mathrm{pt}] $ for this multiplication. Of
particular importance is the class of the affine line, the Lefschetz
motive $\motive = [\mathbb{C}]$, and its formal inverse
$\motive^{-1}$ and square root $\motive^{\frac12}$. It then follows from the scissor relations that the
class of the complex one-torus is $[\complex^*] = [\complex \setminus
0] = \motive -1$.

One has $[\frM ] = [\frS] \,
[\frF]$ whenever $ \frM \xrightarrow{\ \frF \ } \frS$ is a Zariski
locally trivial fibration. For example one can regard the variety $GL(n,\IC)$ as a
locally trivial fibration over the $n$-torus $(\complex^*)^n$; the
fibre $\frF$ is the
stabilizer of a nonzero vector in $\complex^n$ and is isomorphic to
$GL(n-1,\IC) \times \complex^{n-1}$. This implies the recursion
relation $[GL(n,\IC)] = (\motive^n - 1) \, \motive^{n-1}\, [GL(n-1 ,\IC)]$ which is solved by
\begin{equation}
\big[GL(n,\IC) \big] = \motive^{\frac12\, n \, (n-1)} \ \prod_{k=1}^n \, \big(\motive^k
- 1 \big) \ . 
\label{GLnmot}\end{equation}
The above definitions can be extended from varieties to stacks, and in
particular to the moduli spaces of BPS states by formally inverting
the motive $[GL(n,\IC)]$.

For the BPS invariants associated with quivers the relevant
moduli spaces are obtained by cutting the moduli space of quiver
representations by certain matrix relations. These relations follow
from the critical points of a superpotential $W_{\mbf k}$, where the
dimensions of the matrices are encoded as usual in the dimension
vector $\mbf k$ of the quiver representations; in our case
$W_{\mbf k}$ is a cubic polynomial. To the superpotential we associate the
function $f_{\mbf k} := \Tr  W_{\mbf k} : \cM (\sfQ,\mbf k) \rightarrow
\complex$, where $\cM(\sfQ,\mbf k) =\Hom_{\Gamma}
\left( V, Q \otimes V\right)$. Recall that we are
  studying BPS states from a D-brane worldvolume gauge theory perspective in the internal
  space. Equivalently one could take the point of view
  of~\cite{Ooguri:2008yb,Denef:2002ru} who consider the low-energy
  effective $\cN=2$ field theory for the BPS states on $\real^4$. The relations
  which characterize the instanton quiver are then precisely the same as
  the F-term conditions, although of course in our case the use of the
  terminology ``superpotential'' is not really correct. Yet this
  language is useful since in our case we can think of the F-term
  relations as arising from the affine case $X= \complex^3$ upon
  decomposition of the orbifold states into twisted sectors. This
  approach is rather natural when thinking in terms of
  noncommutative instantons. The flat space generalized ADHM equations
  yield the critical points of a superpotential given by the holomorphic
  Chern--Simons action $\Tr Z_1 \,
  [Z_2 , Z_3]$. In the orbifold case one splits the Hilbert
  space of BPS states into twisted sectors corresponding to string states
  transforming as different characters of the orbifold action as in
  (\ref{Hisotop}). One then decomposes the trace into the twisted
  sectors as 
  in (\ref{orbcovcoord}) in the obvious way.

Recall that the arrows in the instanton quivers are
associated with multiplication of the characters of the orbifold
group $\Gamma$; the node structure corresponds to the characters while the
arrow multiplicities follow from the decomposition of the tensor
product with the fundamental representation $Q$ of $\Gamma$, i.e. the
representation $Q=\rho_{r_1}\oplus\rho_{r_2}\oplus\rho_{r_3}$, which contains the information about the orbifold
action on $\complex^3$, determines the arrow structure. This is
equivalent to saying that the individual terms in the superpotential
can be regarded as monomials (actually
invariants) in the characters. Explicitly, the superpotential is realized as a sum over
the character lattice of $\Gamma$ as
\beq
f_{\mbf k}=\Tr W_{\mbf k}= \sum_{r\in\Gammaw}\, \Tr
B_1^{(r+r_2+r_3)}\,\Big( B_2^{(r+r_3)}\, B_3^{(r)}-B_3^{(r+r_2)}\,
  B_2^{(r)} \Big) \ .
\label{TrWkgen}\eeq

The motivic noncommutative Donaldson--Thomas invariant is essentially the
\textit{virtual} motivic class of the critical locus of the function
$f_{\mbf k}$ defined by
\begin{equation}
\big[\NDT_{\mu=0} (\mbf k)\big] = \motive^{\frac12\, \mbf k\cdot\mbf k} \ \frac{\big[\{\dd f_{\mbf k}=0\}\big]}{[G_{\mbf k}]} \ .
\end{equation}
In general, for a function $g : \frM \rightarrow \complex$ the virtual
motive of the locus $\frZ = \{ \dd g = 0 \} $ can be expressed in terms of the motivic vanishing cycle introduced in~\cite{denefloeser} as
\begin{equation}
[\frZ] = - \motive^{-\frac12\, \dim_\IC(\frM)} \ [\varphi_g] \ .
\end{equation}
It is proven in \cite{bbs} that, under favourable conditions including
a certain toric action, the motivic vanishing cycle can be written as
a difference between the motivic classes of the generic fibre and of
the fibre over the origin through
\begin{equation}
[\varphi_g] = \big[g^{-1} (1)\big] - \big[g^{-1} (0)\big] \ .
\end{equation}
In particular this holds in the case of the generalized ADHM equations
for $\complex^3$. As we have argued above, the orbifold case is
obtained by simply decomposing the affine space ADHM equations
according to the twisted sectors as in (\ref{ADHMorb}). We can
therefore assume that the hypotheses of \cite{bbs} hold also in our
case. In particular we may express the Donaldson--Thomas virtual class
in terms of ordinary motivic classes as
\begin{equation}
\big[\NDT_{\mu=0} (\mbf k)\big] = \motive^{\frac12\, \chi_\sfQ (\mbf k , \mbf k)} \
\frac{\big[f_{\mbf k}^{-1} (0)\big] - \big[f_{\mbf
    k}^{-1}(1)\big]}{[G_{\mbf k}]} \ ,
\end{equation}
where $\chi_\sfQ$ is the Euler--Ringel form of the quiver $\sfQ$, i.e. the
bilinear form on $\IN_0^{|\Gamma|}$ given by
\begin{eqnarray}
\chi_\sfQ(\mbf k , \mbf k'\,) = \sum_{r\in\Gammaw} \, k_r\, k_r' -
\sum_{r,s\in\Gammaw} \, a_{sr}^{(1)}\,  k_{r} \, k_s' \ .
\end{eqnarray}
The final missing ingredient is the information about the boundary
conditions which is included in the framing; below we show that this
simply amounts to a minor modification of the above formalism.

We will now adapt the formalism of \cite{nagao,morrison}, which is based on \cite{bbs}, to our instanton quivers. This basically amounts to a minor extension of the formalism of \cite{nagao} to framed quivers, in analogy to what was done in \cite{morrison} for quivers whose superpotential has a linear factor. The key concept is that of a \textit{cut}: a subset of arrows $\sfC \subset \sfQ_1$ together with an $\IN_0$-grading
\begin{equation}
\gr_{\sfC} (a) = \left\{ \  \begin{matrix} 1 \ , && a \in \sfC \ , \\ 0 \
  , && a \notin \sfC \ , \end{matrix} \right.
\end{equation}
such that the superpotential $W_{\mbf k}$ is homogeneous of degree one with respect to
$\gr_{\sfC}$. The degree zero part of $\sfQ$,
which is the quiver $\sfQ_{\sfC}:=(\sfQ_0,\sfQ_1 \setminus\sfC)$, has its own path algebra and
its own category of representations. In particular we have the
representation space $\cM (\sfQ_{\sfC} , \mbf k)$, and its subspace $\cC (\sfQ _{\sfC}
, \mbf k )$ consisting of modules over the path algebra
$\sfA_\sfC= \complex \sfQ_{\sfC} / \langle \partial_a W_{\mbf k} \rangle_{a \in \sfC}$.

To compute the motivic Donaldson--Thomas invariants associated with
the instanton quivers, we first compute the difference $[f_{\mbf
  k}^{-1} (1)] - [f_{\mbf k}^{-1}(0)]$. For this, regard the map
$f_{\mbf k} = \Tr  W_{\mbf k} : \cM (\sfQ , \mbf k) \rightarrow
\complex$ as a fibration. The stratification $\complex = \complex^*
\sqcup {0} $ gives a relation between the fibre over $0$ and the
generic fibre through
\begin{equation} \label{motrel1}
\big[\cM (\sfQ , \mbf k)\big] = \big[f^{-1}_{\mbf k} (0)\big] +
(\motive - 1)\, 
\big[f^{-1}_{\mbf k} (1)\big] \ ,
\end{equation}
where we used $[\complex^*] = \motive -1$. 

Similarly we can find another relation by considering the projection
$\pi : \cM (\sfQ , \mbf k) \rightarrow \cM (\sfQ_{\sfC} , \mbf k)
$. This is a trivial vector bundle of rank
\beq
d(\mbf k) := \dim_\IC \cM
(\sfQ , \mbf k) - \dim_\IC \cM (\sfQ_{\sfC} , \mbf k) =
\chi_{\sfQ_\sfC} (\mbf k , \mbf k)- \chi_\sfQ (\mbf k , \mbf k) \ ,
\eeq
and therefore
$[ \cM (\sfQ , \mbf k) ] = \motive^{d(\mbf k)}\, [\cM (\sfQ_{\sfC} ,
\mbf k) ]$. This allows us to compute the class of $f^{-1}_{\mbf k} (0)$; the locus is a trivial fibration whose base consists now of two strata: $\cC (\sfQ _{\sfC} , \mbf k )$  (since some F-term relations have already been imposed) and $\cM (\sfQ_{\sfC} , \mbf k) \setminus \cC (\sfQ _{\sfC} , \mbf k )$. In the second stratum a linear condition has still to be imposed on the superpotential thus reducing the dimension of the fibre by one. Therefore
\begin{equation} \label{motrel2}
 \big[f^{-1}_{\mbf k} (0)\big] = \motive^{d (\mbf k)} \ \big[\cC (\sfQ
 _{\sfC} , \mbf k )\big] + \motive^{d (\mbf k) - 1} \ \Big( \big[ \cM
 (\sfQ_{\sfC} , \mbf k)\big] -  \big[ \cC (\sfQ _{\sfC} , \mbf k ) \big]
 \Big) \ .
\end{equation}
By merging (\ref{motrel1}) and (\ref{motrel2}) together one finds~\cite[Theorem~4.1]{nagao}
\begin{equation} \label{motrel3}
\big[f_{\mbf k}^{-1} (0)\big] - \big[f_{\mbf k}^{-1}(1)\big] =
\motive^{d (\mbf k)} \ \big[ \cC (\sfQ _{\sfC} , \mbf k )\big] \ .
\end{equation}

We will now extend this construction to framed representations. Recall
that in our formalism (see also \cite{Ooguri:2008yb}) the framing
nodes represent D6-branes which are well-separated in $\real^4$. For
each D6-brane, instanton configurations are constructed starting from
the associated framing node and correspond to monomials in the
representation matrices obtained by acting on a reference vector
associated with the framing node. The framing is specified by a
dimension vector $\mbf N$. In the framed case the definition of the motivic invariant is slightly modified, since the framing factors do not appear in the superpotential. The relevant representation space is now 
\begin{equation}
\cM (\sfQ , \mbf k , \mbf N) = \Hom_{\Gamma} \left( V , Q \otimes V
\right) \ \oplus \ \Hom_{\Gamma} (W , V ) \ .
\end{equation}
Note that the extra factor is just the affine variety $\complex^{\mbf N \cdot \mbf k}$. Therefore if we denote
{ \small
\begin{eqnarray}
\cY_{\mbf N , \mbf k} =  f^{-1}_{\mbf k} (0) \cap \cM (\sfQ , \mbf k ,
\mbf N) \qquad \mbox{and} \qquad 
\cW_{\mbf N , \mbf k} = f^{-1}_{\mbf k} (1) \cap \cM (\sfQ , \mbf k
, \mbf N)
\end{eqnarray} }
then the relation (\ref{motrel3}) implies
\begin{equation} \label{motframed}
[\cY_{\mbf N , \mbf k}] - [\cW_{\mbf N , \mbf k} ]= \motive^{d (\mbf
  k)} \, \motive^{\mbf N \cdot \mbf k}  \ \big[\cC (\sfQ _{\sfC} , \mbf
k ) \big] \ .
\end{equation}

We now closely follow the approach of~\cite{bbs,morrison} to get a
recursion equation for the virtual motivic invariants; in this
approach, we use suitable reduction theorems~\cite{morrison,mozgovoy2}
to express the refined invariants in terms of ordinary classes of
certain reduced quiver representations. For each $\mbf
k$ define the subspace $\cM^{\mbf l} (\sfQ , \mbf k , \mbf N) \subset
\cM(\sfQ , \mbf k , \mbf N)$ spanned by the matrices $B_{\alpha}$ of
the quiver with dimension vector $\mbf l \le \mbf
k$, i.e. $l_r\leq k_r$ for each $r\in\Gammaw$. The definitions (\ref{motframed}) carry over to
{ \small
\begin{eqnarray}
\cY^{\mbf l}_{\mbf N , \mbf k} = f^{-1}_{\mbf k} (0) \cap \cM^{\mbf l}
(\sfQ , \mbf k , \mbf N) \qquad \mbox{and} \qquad
\cW^{\mbf l}_{\mbf N , \mbf k} = f^{-1}_{\mbf k} (1) \cap \cM^{\mbf l}
(\sfQ , \mbf k , \mbf N) \ .
\end{eqnarray} }
In terms of these variables the motivic noncommutative Donaldson--Thomas invariant is
\begin{equation} \label{virtDT}
\big[\NDT_{\mu=0} (\mbf k , \mbf N)\big] = \motive^{-\frac12\, \dim_\IC \cM (\sfQ , \mbf k ,
  \mbf N)} \ \motive^{\frac12\, \mbf k\cdot\mbf k} \ \frac{\big[\cY^{\mbf k}_{\mbf N , \mbf k}\big] -
  \big[\cW^{\mbf k}_{\mbf N , \mbf k}\big]}{[G_{\mbf k}]} \ ,
\end{equation}
with $\dim_\IC \cM (\sfQ , \mbf k , \mbf N) - \frac12\, \mbf
k\cdot\mbf k = - \chi_{\sfQ} (\mbf k ,
\mbf  k) + \mbf  N \cdot \mbf k$. We will now compute this difference
between motivic classes.

Consider first the motive $[\cY^{\mbf l}_{\mbf N , \mbf k}]$. Locally
there is a Zariski fibration over the quiver Grassmannian
\begin{equation}
\cY^{\mbf l}_{\mbf N , \mbf k} \ \longrightarrow \ Gr (\mbf l , \mbf
k):=\prod_{r\in\Gammaw}\, Gr(l_r,k_r)
\end{equation}
which sends an element of $\cY^{\mbf l}_{\mbf N , \mbf k}$ to the array
of subspaces $U_r\subset V_r$ spanned by the matrices $B_{\alpha}$ with dimension
vector $\mbf l$ such that $U_r$ form a subrepresentation of $V$. We have to compute the fibre of this map.

For this,
it is convenient to pick a basis in the vector space $V$ in which the
generalized ADHM matrices can be expressed as
\begin{equation}
B_{\alpha} = \left( \begin{matrix} \tilde{B}_{\alpha} &  \cB_{\alpha}
    \\ 0 & \hat{B}_{\alpha} \end{matrix} \right)  \qquad \mbox{for} \quad
\alpha=1,2,3 \ ,
\end{equation}
whose image on vectors of the form
\begin{equation}
\mbf v=(v_r)_{r\in\Gammaw} \qquad \mbox{with} \quad v_r = \left( \begin{matrix} v_r^{(0)} \\ 0 \end{matrix} \right)
\end{equation}
generate the whole $\mbf l$-dimensional subspace $\cY^{\mbf l}_{\mbf N
  , \mbf k} $; here $v_r^{(0)}$ is an $l_r$-dimensional vector for
each $r\in\Gammaw$.
In this basis the computation simplifies. The Chern--Simons action becomes
\begin{equation}
\Tr W_{\mbf k} (B_{\alpha}) = \Tr W_{\mbf l} (\tilde{B}_{\alpha}) +
\Tr W_{\mbf k- \mbf l} (\hat{B}_{\alpha}) \ .
\end{equation} 
Note that the submatrices $\cB_{\alpha}$ have disappeared; this
implies that they correspond to a trivial fibre of the form
$\complex^{\sum_{r,s\in\Gammaw}\,a_{rs}^{(1)}\, (k_r-l_r)\, l_s
}$. For a vanishing superpotential there are only two possibilities
{ \small
\begin{equation}
\big\{  \Tr W_{\mbf l} (\tilde{B}_{\alpha}) = \Tr W_{\mbf k-\mbf l} (\hat{B}_{\alpha}) =0 \big\}   \qquad \text{or} \qquad
\big\{  \Tr W_{\mbf l} (\tilde{B}_{\alpha}) = -  \Tr W_{\mbf k-\mbf l}
(\hat{B}_{\alpha}) \neq 0 \big\} \ .
\end{equation} }
One can induce a product structure by projecting onto the factors
$(\tilde{B} , \mbf v^{(0)})$ and $(\hat{B})$ (where in the second case
we do \textit{not} consider the span).
This implies that the first stratum corresponds overall to the fiber 
\begin{equation}
\big[\cY^{\mbf l}_{\mbf N , \mbf l}\big] \ \motive^{\sum_{r,s\in\Gammaw}\,
  a_{rs}^{(1)}\, (k_r-l_r)\, l_s} \ [\cY_{\mbf N , \mbf k-\mbf l}] \
\motive^{- \mbf N \cdot (\mbf k-\mbf l)} \ .
\end{equation}
The last factor comes from $[\cY_{\mbf k-\mbf l}] = [\cY_{\mbf N ,
  \mbf k-\mbf l}]
\, \motive^{-\mbf N \cdot (\mbf k-\mbf l)}$, since the framing just
yields a trivial factor when we do not consider the span.
The second stratum is nearly identical. The only difference is that since now both terms are non-vanishing (but equal) there is an additional $\complex^*$-fibration (where the point removed from $\complex$ is precisely the origin where both terms vanish) which gives an extra factor $\motive-1$. Therefore we have
\begin{equation}
(\motive -1) \, \big[\cW^{\mbf l}_{\mbf N , \mbf l}\big] \ \motive^{\sum_{r,s\in\Gammaw}\,
  a_{rs}^{(1)}\, (k_r-l_r)\, l_s} \ [\cY_{\mbf N , \mbf k-\mbf l}] \
\motive^{- \mbf N \cdot (\mbf k-\mbf l)} \ .
\end{equation}

Altogether, by taking into account also the base of the Zariski fibration, this gives
\begin{eqnarray}
\big[\cY^{\mbf l}_{\mbf N , \mbf k}\big] &=& \big[Gr (\mbf l , \mbf k) \big] \ \motive^{\sum_{r,s\in\Gammaw}\,
  a_{rs}^{(1)}\, (k_r-l_r)\, l_s} \ \motive^{- \mbf N \cdot (\mbf
  k-\mbf l)} \ [\cY_{\mbf
  N , \mbf k-\mbf l}] \,  \nonumber \\  && \hspace{2.5cm} \times  \Big( 
\big[\cY^{\mbf l}_{\mbf N , \mbf l} \big] + 
(\motive -1) \, \big[\cW^{\mbf l}_{\mbf N , \mbf l}\big] 
\Big) \ .
\end{eqnarray} 

Next we have to do a similar computation for the motive $[\cW^{\mbf l}_{\mbf N ,
  \mbf k}]$. Again there is a Zariski fibration $[\cW^{\mbf l}_{\mbf N
  , \mbf k}] \rightarrow Gr (\mbf l , \mbf k)$. We can pick a basis as
above. Now however the superpotential has to be set equal to $1$ by definition of $[\cW^{\mbf l}_{\mbf N , \mbf k}]$. Therefore there are three cases
\begin{eqnarray}
\big\{  \Tr W_{\mbf l} (\tilde{B}_{\alpha}) = 1 \ , \ \Tr W_{\mbf
  k-\mbf l} (\hat{B}_{\alpha}) =0 \big\} \ , \nonumber \\[4pt]
\big\{  \Tr W_{\mbf l} (\tilde{B}_{\alpha}) =  0 \ , \ \Tr W_{\mbf
  k-\mbf l} (\hat{B}_{\alpha}) =1 \big\} \ , \nonumber \\[4pt]
\big\{  \Tr W_{\mbf l} (\tilde{B}_{\alpha}) = \zeta \ , \ \Tr W_{\mbf
  k-\mbf l} (\hat{B}_{\alpha}) =1 - \zeta \  , \ \zeta \neq 0,1 \big\} \ .
\end{eqnarray}
The computation proceeds as above; the only difference is the third case
where now the value of $\zeta$ is arbitrary in $\complex \setminus
\{0,1\}$ and therefore gives a factor of $\motive - 2$. These three
cases give respectively the contributions
\begin{eqnarray}
\big[\cW^{\mbf l}_{\mbf N , \mbf l} \big] \ [\cY_{\mbf N , \mbf k-\mbf
  l}] \ \motive^{\sum_{r,s\in\Gammaw}\,
  a_{rs}^{(1)}\, (k_r-l_r)\, l_s} \ \motive^{- \mbf N \cdot (\mbf
  k-\mbf l)} \ , \nonumber\\[4pt]
\big[\cY^{\mbf l}_{\mbf N , \mbf l} \big] \ [\cW_{\mbf N , \mbf k-\mbf
  l}] \ \motive^{\sum_{r,s\in\Gammaw}\,
  a_{rs}^{(1)}\, (k_r-l_r)\, l_s} \ \motive^{- \mbf N \cdot (\mbf
  k-\mbf l)} \ , \nonumber\\[4pt]
(\motive - 2) \, \big[\cW^{\mbf l}_{\mbf N , \mbf l} \big] \
[\cW_{\mbf N , \mbf k-\mbf l}] \ \motive^{\sum_{r,s\in\Gammaw}\,
  a_{rs}^{(1)}\, (k_r-l_r)\, l_s} \ \motive^{- \mbf N \cdot (\mbf
  k-\mbf l)} \ .
\end{eqnarray}

Altogether one therefore finds
{ \small
\begin{eqnarray}
\big[\cW^{\mbf l}_{\mbf N , \mbf k}\big] &=& \big[Gr (\mbf l , \mbf k)\big] \ \motive^{\sum_{r,s\in\Gammaw}\,
  a_{rs}^{(1)}\, (k_r-l_r)\, l_s} \ \motive^{- \mbf N \cdot (\mbf
  k-\mbf l)} \\ 
&& \hspace{-1.5cm} \times \,
 \Big( 
 \big[\cW^{\mbf l}_{\mbf N , \mbf l} \big] \ [\cY_{\mbf N , \mbf k- \mbf
   l}]  + \big[\cY^{\mbf l}_{\mbf N , \mbf l} \big] \ [\cW_{\mbf N , \mbf
   k-\mbf l}] + (\motive - 2)\, \big[\cW^{\mbf l}_{\mbf N , \mbf l}
 \big] \ [\cW_{\mbf N , \mbf k-\mbf l}]
 \Big) \ . \nonumber
\end{eqnarray} }
Finally, to compute the difference $[\cY_{\mbf N , \mbf k}] -
[\cW_{\mbf N , \mbf k}]$, we note that each term stratifies as
$\cY_{\mbf N , \mbf k} = \bigsqcup_{\mbf l \le \mbf k}\, \cY^{\mbf
  l}_{\mbf N , \mbf k}$ and
$\cW_{\mbf N , \mbf k} = \bigsqcup_{\mbf l \le \mbf k} \, \cW^{\mbf
  l}_{\mbf N , \mbf k}$. We can then go back to the expression (\ref{motframed}) to write
{ \small
\begin{eqnarray}
\motive^{d (\mbf k)} \, \motive^{\mbf N \cdot \mbf k} \ \big[\cC (\sfQ
_{\sfC} , \mbf k ) \big]  \hspace{-2cm} && \\[4pt] \nonumber  &=& \sum_{\mbf l \le \mbf k} \, \Big( \big[
\cY^{\mbf l}_{\mbf N , \mbf k}\big] - \big[ \cW^{\mbf l}_{\mbf N ,
  \mbf k}\big] \Big) \\[4pt]
&=& \sum_{\mbf l \le \mbf k} \ \big[Gr (\mbf l , \mbf k)\big] \  \motive^{\sum_{r,s\in\Gammaw}\,
  a_{rs}^{(1)}\, (k_r-l_r)\, l_s} \ \motive^{- \mbf N \cdot (\mbf
  k-\mbf l)}\, \Big( 
\big[\cY^{\mbf l}_{\mbf N , \mbf l}\big] \  [\cY_{\mbf N , \mbf k-\mbf
  l}] \nonumber \\ && \ +\, 
(\motive -1) \, \big[\cW^{\mbf l}_{\mbf N , \mbf l}\big] \ [\cY_{\mbf
  N , \mbf k-\mbf l}]
- \big[\cW^{\mbf l}_{\mbf N , \mbf l} \big] \ [\cY_{\mbf N , \mbf
  k-\mbf l}]  - \big[\cY^{\mbf l}_{\mbf N , \mbf l} \big] \ [\cW_{\mbf N
  , \mbf k-\mbf l}]  \nonumber \\ && \ -\, (\motive - 2)\,
\big[\cW^{\mbf l}_{\mbf N , \mbf l} \big] \ [\cW_{\mbf N , \mbf k-
  \mbf l}] 
\Big) \nonumber \\[4pt]
&=& \sum_{\mbf l \le \mbf k} \ \big[Gr (\mbf l , \mbf k)\big] \  \motive^{\sum_{r,s\in\Gammaw}\,
  a_{rs}^{(1)}\, (k_r-l_r)\, l_s} \ \motive^{- \mbf N \cdot (\mbf
  k-\mbf l)} \nonumber \\ && \qquad \times\, \Big( 
\big[\cY^{\mbf l}_{\mbf N , \mbf l}\big] - \big[\cW^{\mbf l}_{\mbf N , \mbf l} \big] \Big)\, \Big( 
[\cY_{\mbf N , \mbf k-\mbf l}]  - 
 [\cW_{\mbf N , \mbf k-\mbf l}] 
\Big) \ . \nonumber 
\end{eqnarray} }
Now if we use (\ref{motframed}) again as well as the definition
(\ref{virtDT}) then we find
{ \small
\begin{eqnarray}
\motive^{d (\mbf k)}\, \motive^{\mbf N \cdot \mbf k}  \ \big[\cC (\sfQ _{\sfC} , \mbf k )\big] 
 \hspace{-2cm} && \\[4pt] \nonumber 
&=&\sum_{\mbf l \le \mbf k} \ \big[Gr (\mbf l , \mbf k)\big] \  \motive^{\sum_{r,s\in\Gammaw}\,
  a_{rs}^{(1)}\, (k_r-l_r)\, l_s} \ \motive^{- \mbf N \cdot (\mbf
  k-\mbf l)} \ \motive^{d(\mbf k-\mbf l)} \, \motive^{\mbf N \cdot (\mbf
  k-\mbf l)}
\nonumber \\ && \qquad \times\, \big[\cC (\sfQ _{\sfC} , \mbf k-\mbf l)
\big] \ [G_{\mbf l}] \ \big[\NDT_{\mu=0} (\mbf l , \mbf N)\big] \
\motive^{\frac{1}{2}\, ( - \chi_{\sfQ} (\mbf l , \mbf l)+ \mbf l \cdot
  \mbf N )} \ . \nonumber
\end{eqnarray} }

We can simplify this last relation by expanding the class of the
quiver Grassmanian. One has
\begin{eqnarray}
\big[Gr (\mbf l , \mbf k )\big] = \frac{[G_{\mbf k}]}{[G_{\mbf l}]
  \ [G_{\mbf k-\mbf l}] \big[ \IC^{\mbf l \cdot( \mbf k-\mbf l)} \big]} 
= \frac{[G_{\mbf k}]}{[G_{\mbf l}] \ [G_{\mbf k-\mbf l}] } \
\motive^{-\mbf l \cdot (\mbf k - \mbf l)}
\end{eqnarray}
and by (\ref{GLnmot}) the relevant classes are of the form
\begin{equation}
[G_{\mbf k}] = \prod_{r \in \Gammaw}\, \big[GL(k_r,\IC) \big] =
\prod_{r\in \Gammaw} \ \motive^{\frac{1}{2}\, k_r\, (k_r-1)} \
\prod_{j_r=1}^{k_r}\, \big(\motive^{j_r} -1 \big) \ .
\end{equation}

Putting everything together we arrive at a recursion relation for the
virtual motivic noncommutative invariants given in terms of motives of moduli of
$\sfA_\sfC$-modules by
\begin{eqnarray} \label{motivicrec}
\motive^{d (\mbf k)}\, \motive^{\mbf N \cdot \mbf k}  \ \big[\cC (\sfQ
_{\sfC} , \mbf k )\big] 
 \hspace{-2cm} && \\[4pt] \nonumber 
&=& \sum_{\mbf l \le \mbf k} \ \big[\cC (\sfQ
_{\sfC} , \mbf k-\mbf l) \big] \
\prod_{r\in \Gammaw} \ \motive^{\frac{1}{2}\, (2 k_r -l_r -1)\, l_r} \
\prod_{j_r = k_r - l_r +1}^{k_r} \ \big(\motive^{j_r} - 1 \big) \nonumber
\\ &&
 \times \,
\motive^{-\frac{1}{2}\, \chi_{\sfQ} (\mbf l , \mbf l) - \chi_{\sfQ}
  (\mbf k-\mbf l , \mbf l)} \ \motive^{d (\mbf k-\mbf l)} \
\motive^{-\frac{1}{2}\, \mbf N \cdot (\mbf k-\mbf l)} \, \big[\NDT_{\mu=0}
(\mbf l , \mbf N)\big] \ . \nonumber
\end{eqnarray} 

Let us now study a couple of explicit examples and their associated
combinatorial problems. Consider again the $ \complex^3 / \zed_3$
singularity. In this case the superpotential has the form
\begin{eqnarray}
f_{\mbf k} &=& \Tr \Big( B^{(2)}_3\, \big( B^{(1)}_1\, B^{(0)}_2 -
B^{(1)}_2\, B^{(0)}_1 \big) + B^{(2)}_1 \, \big( B^{(1)}_2\, B^{(0)}_3
- B^{(1)}_3\, B^{(0)}_2 \big) \nonumber
\\[4pt] && \hspace{4cm}
+ B^{(2)}_2\, \big( B^{(1)}_3\,
B^{(0)}_1 - B^{(1)}_1\, B^{(0)}_3 \big) \Big) \, . 
\end{eqnarray} 
The matrices $B^{(2)}_{\alpha}$, $\alpha=1,2,3$ form a cut. After
removing them from (\ref{quiverC3Z3}), the quiver $\sfQ_\sfC$ takes the form
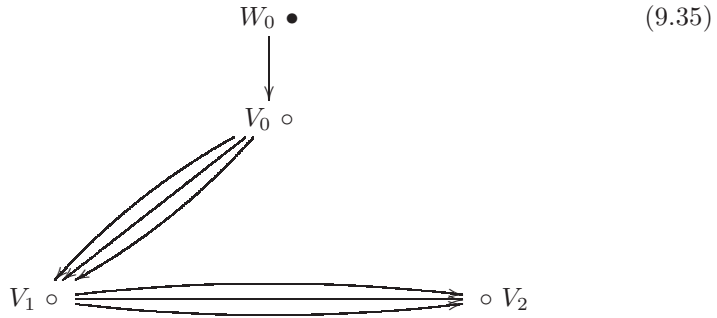
\begin{equation}
\vspace{4pt}
\begin{xy}
\xymatrix@C=20mm{ & W_0 \ \bullet \ar[d] & \\
& \ V_0 \ \circ \ \ar@/^/[ddl] \ar@/_0.5pc/[ddl] \ar@//[ddl]  & \\
& & \\
V_1 \ \circ \ \ar@//[rr] \ar@/^/[rr]  \ar@/_/[rr]   & &  \ \circ \ V_2 }
\end{xy}
\vspace{4pt}
\label{quiverP2}\end{equation}
and the combinatorial problem
that one is left with is the enumeration of sets of matrices obeying the reduced coupled equations
{ \small
\begin{eqnarray}
B^{(1)}_1 B^{(0)}_2 = B^{(1)}_2 B^{(0)}_1 \ , \ 
 B^{(1)}_2 B^{(0)}_3 = B^{(1)}_3 B^{(0)}_2 \  \ \mbox{and} \ \
 B^{(1)}_3 B^{(0)}_1 = B^{(1)}_1 B^{(0)}_3 \ .
\end{eqnarray} }
This is the (framed) Beilinson quiver for $\PP^2$; the representations
of this quiver correspond to (framed) coherent sheaves on the
projective plane which classify BPS states of D4-branes wrapping $\PP^2$ in the
large radius limit. Thus while the
recursion relation (\ref{motivicrec}) is still a difficult
problem, it reduces the computation of the motivic noncommutative invariants to
something considerably easier since in this case the path algebra of
the ``cut'' quiver $\sfQ_\sfC$ has no oriented cycles. 
Geometrically one can understand this result by saying that the
motivic invariants near the orbifold point are
captured by retraction to the zero section of the crepant resolution $X= \mathrm{Hilb}^{\zed_3}
(\complex^3)\cong \cO_{\PP^2}(-3)$ of the
$\IC^3/\zed_3$ singularity (although of course this is not literally
true since the stability conditions are still associated with the
noncommutative crepant resolution). This is just another manifestation
of the McKay correspondence. In~\cite[Section~8.4]{Cirafici:2010bd}
an analogous relation between the derived categories of coherent sheaves on $\PP^2$ and on $X$ is described; it demonstrates that certain holomorphic objects near the orbifold point come from representations of the Beilinson quiver for $\PP^2$.

Next we consider the example of the $\complex^3 / \zed_2 \times \zed_2$
orbifold. The corresponding crepant resolution is semi-small, so for
each arrow of the $\complex^3 / \zed_2 \times \zed_2$ quiver connecting a pair of
nodes there is a dual arrow in the opposite direction. The quiver
$\sfQ$ is
\begin{equation}
\vspace{4pt}
\begin{xy}
\xymatrix@C=20mm{
& \ V_0 \ \circ \ \ar@/^/[ddr] \ar@/_1pc/[ddl]  \ar@/^/[d]& \\
& \ V_3 \ \circ \ \ar@/^/[u] \ar@/^/[dl] \ar@/^/[dr]& \\
V_1 \ \circ \ \ar@/^/[uur] \ar@/^/[ur] \ar@//[rr] & & \ \circ \ V_2  \ar@/_1pc/[uul] \ar@/^/[ul] \ar@/^/[ll]
}
\end{xy}
\end{equation}
and a possible choice of cut $\sfQ_\sfC$ is given by
\begin{equation}
\begin{xy}
\xymatrix@C=20mm{
& \ V_0 \ \circ \ \ar@/^/[ddr] \ar@/_1pc/[ddl]  \ar@/^/[d]& \\
& \ V_3 \ \circ \  \ar@/^/[dr]& \\
V_1 \ \circ \ \ar@/^/[uur] \ar@/^/[ur] \ar@//[rr] & & \ \circ \ V_2 \ar@/^/[ul] 
}
\end{xy}
\vspace{4pt}
\end{equation}
If we introduce the motive
\begin{equation}
\mathbb{V}_{\mbf k} := \motive^{d (\mbf k)} \, \frac{\big[\cC (\sfQ_{\sfC} , \mbf k)\big]}{[G_{\mbf k}]}
\label{Vmotive}\end{equation}
then the recursion relation (\ref{motivicrec}) becomes
\begin{equation}
\motive^{\frac12\, \mbf N \cdot \mbf k} \, \mathbb{V}_{\mbf k} =
\sum_{\mbf l \le \mbf k} \ \motive^{-\frac{1}{2}\, (\mbf k-\mbf l)
  \cdot \mbf N} \ \mathbb{V}_{\mbf k-\mbf l} \, \motive^{\chi_{\sfQ}
  (\mbf k-\mbf l,\mbf l)} \, \motive^{-\frac{1}{2}\, \chi_{\sfQ} (\mbf
  l , \mbf l )} \ \big[\NDT_{\mu=0} (\mbf l , \mbf N)\big] \ .
\end{equation}
Since the Euler--Ringel form $\chi_\sfQ$ is symmetric in this case, it satisfies
\begin{equation}
-\chi_{\sfQ} (\mbf k-\mbf l , \mbf l) - \mbox{$\frac{1}{2}$}\,
\chi_{\sfQ} (\mbf l , \mbf l) = \mbox{$\frac{1}{2}$}\, \chi_{\sfQ}
(\mbf k-\mbf l , \mbf k-\mbf l) - \mbox{$\frac{1}{2}$}\, \chi_{\sfQ}
(\mbf k , \mbf k ) \ .
\end{equation}
Introducing variables $\mbf p=(p_r)_{r\in\Gammaw}$ weighting the (non-framing) nodes of the quiver, we find the relation
\begin{eqnarray}
\sum_{\mbf k\geq\mbf 0}\, \motive^{\frac{1}{2}\, \chi_{\sfQ} (\mbf k , \mbf
  k)}\, \motive^{\frac{1}{2}\, \mbf N \cdot \mbf k}\, \mathbb{V}_{\mbf
  k} \ \mbf{p}^{\mbf k} \hspace{-3.6cm} && \\[4pt] \nonumber &=& \sum_{\mbf k\geq\mbf 0} \ \sum_{\mbf l \le \mbf k} \,
\motive^{- \frac{1}{2}\, \mbf N \cdot (\mbf k-\mbf l)} \
\mathbb{V}_{\mbf k-\mbf l} \ \motive^{\frac{1}{2}\, \chi_{\sfQ} (\mbf
  k-\mbf l , \mbf k-\mbf l)} \ \big[\NDT_{\mu=0} (\mbf l , \mbf N ) \big] \
\mbf{p}^{\mbf k-\mbf l} \ \mbf{p}^{\mbf l} \ .
\end{eqnarray} 
As in \cite{bbs} the sums decouple and one arrives finally at the
motivic BPS partition function
{ \small
\begin{equation}
\mathcal{Z}_{\IC^3/\zed_2 \times \zed_2}^{\rm mot}(\mbf N) :=
\sum_{\mbf k\geq\mbf 0}\, \big[\NDT_{\mu=0} (\mbf k , \mbf N)\big] \ \mbf{p}^{\mbf k} =
\frac{\sum\limits_{\mbf k\geq\mbf 0}\, \motive^{\frac{1}{2}\, \chi_{\sfQ} (\mbf
    k , \mbf k)} \, \motive^{\frac{1}{2}\, \mbf N \cdot \mbf k}\,
  \mathbb{V}_{\mbf k} \ \mbf{p}^{\mbf k}}{\sum\limits_{\mbf m\geq\mbf 0}\,
  \motive^{\frac{1}{2}\, \chi_{\sfQ} (\mbf m , \mbf m )}\,
  \motive^{-\frac{1}{2} \, \mbf N \cdot \mbf m}\, \mathbb{V}_{\mbf m }
  \ \mbf{p}^{\mbf{m}}} \ .
\label{ZmotZ2Z2}\end{equation} }
Motivic partition functions such as (\ref{ZmotZ2Z2}) should be
compared with their refined counterparts (\ref{Zreforb}) under the
identifications (\ref{NAinst}) and the \linebreak refined/motivic correspondence
$\lambda=\motive^{\frac12}$~\cite{Dimofte:2009bv,MMNSz}.

It would be interesting to find a closed form for the
motivic partition functions in this case, precisely as is the situation for the ordinary
(unrefined) invariants. Presumably such formulas come from suitable
generalizations of the partition function~\cite[Proposition~1.1]{bbs}
\beq
\sum_{k=0}^\infty \ \frac{[C_k]}{\big[GL(k,\complex)\big]} \ q^k =
  \prod_{n=1}^\infty \ \prod_{m=0}^\infty\, \big(1-\motive^{1-m} \ q^n
  \big)^{-1}
\label{partCk}\eeq
for the class of the variety $C_k$ of commuting pairs of $k\times k$ complex
matrices, which could be used to perform the sums involving the motives
(\ref{Vmotive}). Analogous closed expressions are computed
in~\cite{morrison,mozgovoy2} for the abelian orbifolds
$\complex^2/\zed_n\times\complex$ of type $A_{n-1}$ by retraction to $\complex^2/\zed_n$. The generalized McKay quiver
is again symmetric, and now contains a loop at each node~\cite{Cirafici:2010bd}. In this case the recursion and
reduction formulas can be evaluated explicitly via two natural cuts:
firstly by taking $\sfC$ to be the set of vertex loops so that
$\cC(\sfQ_\sfC,\mbf k)$ consists of modules over the preprojective
algebra of the standard affine McKay quiver for the four-dimensional
singularity $\complex^2/\zed_n$ and generalizing the partition
function (\ref{partCk}), and
then further cutting with the collection of all dual arrows $\sfC'$ so
that $(\sfQ_\sfC)_{\sfC'}$ coincides itself with the affine Dynkin
diagram of type $\hat A_{n-1}$. Thus
in this case the process of dimensional reduction reduces the problem
to that of representations of the simply-laced extended Dynkin quiver
of type $\hat A_{n-1}$, analogously to the reduction to the Beilinson
quiver for $\PP^2$ that we encountered above.

\section{Wall-crossing formulas from McKay data}

To the generators of the charge lattice $\Lambda$ we associate sets of invertible
operators $\{ \sfX_{I} \}$ and their inverses which includes operators $\{ \sfX_{r} \}$
corresponding to the irreducible representations $\rho_r $,
$r\in\widehat\Gamma$ and $\{ \sfX_{\infty} \}$ corresponding to the
framing nodes. These operators generate the quantum torus algebra
associated with the basis of fractional branes; this is the
associative noncommutative algebra over $\complex$ defined by the relations
\begin{equation}
\sfX_{I} \, \sfX_{J}  =  \lambda^{2\left( \gamma_I , \gamma_J \right)}\ \sfX_{J}\, \sfX_{I}
\end{equation}
where $\lambda$ is the spin weighting parameter introduced in
(\ref{qlambdaweights}). Similarly, for the lattice $\widehat{\Lambda}$ we can define a quantum torus $\torus_{\widehat{\Lambda}}^{*}$ by the same set of operators $ \{ \sfX_{I} \} = \{ \sfX_{r} , \sfX_{\infty} \}$ but with the new commutation relations
\begin{equation}
\sfX_{I}\, \sfX_{J} = \lambda^{2\langle \gamma_I , \gamma_J \rangle} \ \sfX_{J}\, \sfX_{I} \ .
\end{equation}
For an instanton quiver with trivial framing the non-trivial relations
are generally
\begin{eqnarray}
\sfX_{r} \, \sfX_{s} = \lambda^{2 a^{(2)}_{rs} - 2 a^{(1)}_{rs}}\ \sfX_{s} \, \sfX_{r} \qquad \mbox{and} \qquad
\sfX_{\infty} \, \sfX_{0} = \lambda^2 \ \sfX_{0} \, \sfX_{\infty} \ .
\end{eqnarray}
Now all the information about the quantum torus algebra is encoded in
the group theory associated with the orbifold singularity. In fact,
the generators $\sfX_r$ are the ``quantum'' analogs of the K-theory
generators $\cS_r$, since their algebra is determined entirely by the
intersection pairing on $K^c (X)$. For an arbitrary charge vector
$\gamma=\sum_{r\in\Gammaw}\, g_r\,\gamma_r\in\widehat{\Lambda}$, $g_r\in\IZ$ the
corresponding operator is
\beq
\sfX_\gamma=\lambda^{-\sum_{r< s}\, g_r\, g_s\,(a^{(2)}_{rs} -
  a^{(1)}_{rs})} \ \prod_{r\in\Gammaw}^{\curvearrowright} \,
\sfX_r^{g_r} \ ,
\eeq
which can be extended to include also the framing nodes; here the
product is taken in increasing order with respect to a suitable
lexicographic ordering on the character lattice. Since the
intersection form $\langle-,-\rangle$ on $\widehat{\Lambda}$ is
non-degenerate, there are no central elements in the quantum torus
algebra when the deformation parameter $\lambda$ is not a root of unity.

We can now put all of our ingredients together to construct a quantum
monodromy operator~\cite{Cecotti:2010fi} associated with the virtual
instanton quiver as
\begin{equation}
\sfM (\lambda) = \prod_{\theta_{\rho}}^{\curvearrowright} \, \Psi
\left(  \lambda^{2 s_{\rho}}\, \sfX_{\rho} ; \lambda
\right)^{\Omega^{\rm ref}_{2s_\rho}(\rho)} \ ,
\label{monop}\end{equation}
where the product is over all the (ordered) framed representations of the
orbifold group $\Gamma$. The quantum dilogarithm function $\Psi (x ;
\lambda)$ is defined
as the Pochhammer symbol $(-\lambda\,x;\lambda^2)_\infty$, i.e. the (convergent) infinite product
\begin{equation}
\Psi (x ; \lambda) = \prod_{n=0}^{\infty}\, \big( 1 + \lambda^{2 n+1} \, x \big) \ .
\end{equation}
Here $s_{\rho}$ is the spin content of the BPS states associated with
the (generally reducible) representation $\rho$, and
$\Omega^{\rm ref}(\rho;\lambda) = \sum_n\, \Omega^{\rm ref}_n(\rho)\, (-\lambda)^n \in\zed(\lambda)$ is the corresponding refined
index of states; the product over BPS states is ordered according to increasing
central charge phases $\theta_{\rho}$ of the states. We conjecture that the conjugacy class of this operator, in analogy
with the construction of \cite{Cecotti:2010fi}, is constant upon
crossing a wall of marginal stability where the phases $\theta_\rho$
and $\theta_{\rho'}$ for two linearly independent $\rho$, $\rho'$
become aligned. We will discuss later on how to
rephrase this operation, and hence the combinatorics of the
wall-crossing jumps, in purely group theory terms via the theory of cluster algebras associated with the quiver. All the data involved in this operator, including the commutation relations, are purely group theoretical and explicitly known once the orbifold singularity is chosen, \textit{except} for the BPS multiplicities, the spins of the states, and their central charges.  

We can however take a further step and try to identify the central
charges and their phases, at least in the context of stacky gauge
theories; at this stage though it is far from clear that this ordering
of states coincides with the ordering of string theory BPS states, due
to the caveats already expressed in~\cite{Cirafici:2010bd}. Motivated
by considerations of $\Pi$-stability, it is natural to define a
quiver
central charge function $\sfZ_{\omega,B}:\Lambda \to
\complex$ via the McKay correspondence as the total
charge of a D6--D4--D2--D0 bound state on $X$; it is linear on
the charge lattice $\Lambda$ and is given by
integrating the (twisted) Mukai vector of the corresponding coherent sheaf $\cE$
to get
\begin{equation} \label{central}
\sfZ_{\omega,B} (\cE) = \int_X \, \e^{-B -\ii \omega} \wedge \ch (\cE) \wedge \sqrt{{\rm Todd} (X)} \ .
\end{equation}
This definition is intended with the following prescription. The Chern
character of a torsion-free sheaf $\cE$ on $X$ is computed as in
(\ref{chE}) while the Todd class, as well as the complexified K\"ahler
moduli $B+\ii\omega$, are expressed via the basis of $H^{\sharp}
(X,\complex)$ given by the Chern classes of the set of tautological
bundles (\ref{cRdecomp}). This prescription can be carried out quite
explicitly in several cases by using the calculations
of~\cite{Cirafici:2010bd}. Of course the explicit result will depend
on the particular orbifold singularity; for example with trivial boundary
conditions at infinity one generically finds for
the real and imaginary parts
{ \small
\begin{eqnarray}  \label{ReZE} 
\Re\, \sfZ_{\omega,B} (\cE) \hspace{-1.8cm} && \\ \nonumber &=& \tch_0(\cE)\,\Big(\,
\frac1{24}\, \big(c_1\, c_2-2c_1^3\big)-\frac{c_1}4\,
\sum_{m\in\Gammaw}\, \varsigma_m\, c_2(\cV_m) \, \Big)\\
&& +\,
\sum_{r,s\in\Gammaw}\, k_s\, \bigg[\langle\gamma_s,\gamma_r\rangle\,
\bigg(\, \tch_1(\cR_r)\, \Big(\, \frac1{24}\, \big(2c_2-c_1^2\big)
-\frac12\, \sum_{m\in\Gammaw}\, \varsigma_m\, c_2(\cV_m)\, \Big)
\nonumber \\ && \qquad \qquad \qquad \qquad \qquad +\,
\frac{\ch_1^{(1)}}2 \, \tch_2(\cR_r)
+\tch_3(\cR_r)\bigg) + 6\delta_{rs}\, \tch_3^{(1)} \nonumber \\ &&
+\, \big( (\gamma_s,\gamma_r)-3\delta_{rs}\big)\, \Big(
-\frac{c_1\, \ch_1^{(1)}}2\, \tch_1(\cR_r)+\ch_1^{(1)}\ \tch_2(\cR_r)+
\ch_2^{(1)}\ \tch_1(\cR_r)\Big) \bigg] \ , \nonumber \\[4pt] \Im \,
\sfZ_{\omega,B} (\cE) 
 \label{ImZE} 
 \hspace{-1.8cm} && \\ \nonumber
&=& -\frac{\omega^3}6\, \tch_0(\cE) 
+\sum_{n\in\Gammaw}\, \varphi_n\, c_1(\cR_n)\, \bigg[\,
\frac1{24}\, \tch_0(\cE)\,\big(2c_2-c_1^2\big)\\ && 
+\, \sum_{r,s\in\Gammaw}\, k_s\,
\bigg(\langle\gamma_s,\gamma_r\rangle\, \Big( \, \frac{c_1}2\,
\tch_1(\cR_r)+\tch_2(\cR_r) \, \Big) 
-(\gamma_s,\gamma_r)\, \ch_1^{(1)}\
\tch_1(\cR_r)\bigg) \, \bigg] \ . \nonumber
\end{eqnarray} }
Here $\tch$ denotes the twisted Chern character
$\tch(\cE)=\e^{-B}\wedge\ch(\cE)$, \linebreak $c_i:=c_i(X)$ are the Chern classes
of the tangent bundle of $X$, and $\ch_i^{(1)}:=
\ch_i(\cO_{X}(1))$.

The central charge is thus determined by twisted
intersection indices, whose
dependence on the geometric moduli is fixed by the McKay
correspondence. Note that the central charge computed according to
this prescription is linear in the D-brane charges $N=\ch_0(\cE)$ and
$k_s$, as expected. Furthermore, it is consistent
with the known definitions of stability parameters used to analyse BPS
states associated with quivers, such as $\theta$-stability. In
particular, the usual slope function $\mu=\mu_{\omega,B}(\cE)$ defined by the imaginary part
(\ref{ImZE}) depends on both stable and virtual instanton quiver
lattice pairings between fractional D0-branes, as well as the D2-brane
(but not D4-brane) chemical potentials. For a polarization
$\omega\to\infty$, these would determine stability
conditions near the large radius point in the K\"ahler moduli
space. In this region our wall-crossing conjecture may thus be refined by saying
that the quantum monodromy $\sfM (\lambda)$ is constant under
variation of $\sfZ_{\omega,B}\in\{r\, \e^{\ii\theta} \ | \ r>0 \ , \ 0<\theta\leq1\}$. Again, having fixed once and for all the orbifold singularity, all the
data in (\ref{central})--(\ref{ImZE}) are expressed in terms of representation
theory and the physical properties of the instanton one is
considering, i.e. the choice of boundary condition and the
specification via the plane partitions $\pi_{l,r}$ of the instanton
configurations. Moreover, the BPS degeneracies $\Omega^{\rm ref}(\mbf k,\mbf N ;\lambda)$ are computed in the noncommutative crepant resolution chamber
by the formalism of Section~\ref{sec:refined} where they can be
identified with the refined invariants $\NDT^{\rm ref}_{\mu=0}
(\mbf k , \mbf N; \lambda)$. Note that the stability condition $\mu=0$
is equivalent to the vanishing of (\ref{ImZE}) in the large radius chamber.

In the ``classical'' limit, where the quantum (or motivic) parameter
$\lambda\to1$, this formalism rewrites the wall-crossing formula which
was derived by Kontsevich and Soibelman within a local approach based
on Lie algebras. In this case instead of the operators $\sfX_{I}$
considered before we introduce the Lie algebra generated by the
elements $\sfe_{I}:= \lim_{\lambda\to1}\, \sfX_I\big/\big(\lambda^2-1\big)$ which obey the commutation relations
\begin{equation} \label{KSalgebra}
[ \sfe_{I} , \sfe_{J} ] = (-1)^{\langle \gamma_I , \gamma_J \rangle} \, \langle \gamma_I , \gamma_J \rangle \ \sfe_{\gamma_I+\gamma_J} \ ,
\end{equation}
where $\sfe_{\gamma_I+\gamma_J}$ is the generator associated with the
charge $\gamma_I+\gamma_J$. In particular we can define the group-like
operator 
\begin{equation} \label{KSop}
\mathcal{U}_{\rho} = \exp \Big(\, \sum_{n=1}^{\infty}\, \frac{\sfe_{n\, \rho}}{n^2}\, \Big)
\end{equation}
which generates symplectomorphisms of the classical complex torus
$\widehat{\Lambda}\,^{\vee}\otimes \IC^*$. 
Using this operator we can formulate a representation theory version of the Kontsevich--Soibelman conjecture: the operator
\begin{equation} \label{WCop}
\mathcal{K} = \prod_{\theta_\rho}^{\curvearrowright} \, \mathcal{U}_{\rho}^{\Omega (\rho)}
\end{equation}
is constant upon crossing walls of marginal stability. Here the
operators are ordered with the phase of the central charge as
determined in (\ref{central})--(\ref{ImZE}) increasing. The degeneracy
$\Omega(\rho)\in\zed$ represents the index of BPS states with charge given by
the (reducible) representation $\rho$ of $\Gamma$ (recall that in our
notation this may include framing nodes ${\bullet}$ which represent
the D6-brane charge). Once again, except for the degeneracies
$\Omega(\rho)$, all the information entering into the operator
(\ref{WCop}) is completely determined via the McKay correspondence by
the (virtual) instanton quiver and therefore by the orbifold
singularity. Moreover, by using the formalism developed in
\cite{Cirafici:2010bd} the BPS degeneracies $\Omega(\mbf k,\mbf N)$ are known at least in one
chamber, the one corresponding to the noncommutative crepant
resolution where they can be identified with the invariants $\NDT_{\mu=0} (\mbf k , \mbf N)$. From this conjecture one can naturally deduce
wall-crossing behaviours for the noncommutative Donaldson--Thomas
invariants from the associated Lie algebra elements
$\DT_{\mu=0}(\mbf k)\, \sfe_{\mbf k}$ by using
(\ref{NCDTgenrel}), (\ref{quivergeneralized}) and the combinatorial formula
of~\cite[Theorem~7.17]{joycesong}; in particular, for semi-small
crepant resolutions the invariants themselves are independent of the
slope stability conditions.

The connection between the quantum and classical monodromy operators
(\ref{monop}) and (\ref{WCop}) can be elegantly formulated by
similarly relating the motivic wall-crossing formulas with the McKay
correspondence. In this case the motivic quantum torus is generated by
elements $\widehat{\sfe}_I$ associated with the irreducible
representations of the orbifold group, and obeying the twisted
multiplication rule
\begin{equation}
\widehat{\sfe}_I\, \widehat{\sfe}_J = \motive^{\frac{1}{2}\, \langle
  \gamma_I , \gamma_J \rangle} \ \widehat{\sfe}_{\gamma_I+\gamma_J} \ .
\end{equation}
The quantum analog of the operator $\mathcal{U}_{\rho}$ introduced in
(\ref{KSop}) can be now defined via the motivic quantum dilogarithm
\begin{equation}
\mathcal{U}_{\rho} \ \longmapsto \ \Psi \big(\, \widehat{\sfe}_{\rho} \,;\,
\motive^{\frac12}\,\big) \ .
\end{equation}
The motivic wall-crossing formula is the statement that the quantum
monodromy associated with these operators, i.e. the appropriately
ordered product, is invariant and chamber independent. The
corresponding behaviours of the motivic Donaldson--Thomas invariants
are deduced via the element $\mathbb{NC}_{\mu=0}(\mbf k,\mbf N):= \big[\NDT_{\mu=0} (\mbf
k, \mbf N)\big]\, \widehat{\sfe}_{\mbf k}$ of the motivic quantum torus algebra~\cite{mozgovoy1}.

\section{Cluster algebras and the McKay correspondence}

Given a 2-acyclic quiver we can construct its \textit{cluster algebra}
and the associated quantum algebras~\cite{fock,Cecotti:2010fi}
(see~\cite{keller} for a review). This is
the algebra generated by the operators $\sfX_I$ and their (quantum)
mutations by monodromy operators; a
related approach has been pursued in~\cite{nagaocluster}. Roughly
speaking if two quivers are related by a mutation, then they
correspond to different BPS chambers separated by a wall of marginal
stability and
their noncommutative Donaldson--Thomas invariants (more precisely the
motivic algebra elements $\mathbb{NC}_{\mu=0}(\mbf k,\mbf N)$) are linked by the
composition of a monomial transformation with conjugation by a quantum
dilogarithm operator~\cite{Kontsevich:2009xt}. The cluster algebras
can be interpreted as K-theory invariants associated with categories
of quiver representations, and are in this way related to the
categorification described in Section~\ref{sec:motivic}.

Let $\sfQ$ be a 2-acyclic quiver associated with a lattice of charges
$\widehat{\Lambda}$ and the quantum torus
$\torus_{\widehat{\Lambda}}^{*}$. This quiver is obtained from the
virtual instanton quiver associated with an orbifold singularity via the McKay correspondence as we have explained in the previous sections. A mutation of this quiver at the node $K$, $\mut_K(\sfQ)$, is obtained by reversing all the arrows incident to the node $K$ and modifying the arrows between $I$ and $J \neq I$ for all the other vertices different from $K$, according to the rule depicted schematically as
\begin{equation} \label{mutation}
\vspace{4pt}
\begin{tabular}{|c|c|}
\hline
$\sfQ$ & $ \mut_K (\sfQ)$ \\ \hline
$\begin{xy} \xymatrix@C=6mm{ I \ar[rr]^l \ar[dr]_m & & J \\ & K \ar[ur]_n &  } \end{xy} $
&
$\begin{xy} \xymatrix@C=6mm{ I \ar[rr]^{l+m\, n} & & J \ar[dl]^n \\ & K \ar[ul]^m &  } \end{xy} $
\\ \hline
$\begin{xy} \xymatrix@C=6mm{ I \ar[rr]^l & & J \ar[dl]^n \\ & K  \ar[ul]^m &  } \end{xy} $
&
$\begin{xy} \xymatrix@C=6mm{ I \ar[rr]^{l-m\, n} \ar[dr]_m & & J \\ & K \ar[ur]_n &  } \end{xy} $
\\
\hline
\end{tabular}
\vspace{4pt}
\end{equation}
where $l$, $m$ and $n$ are integers, and the notation $\begin{xy} \xymatrix@C=6mm{I \ar[r]^l &J } \end{xy} $ denotes $l$ arrows from $I$ to $J$ if $l\geq0$ and $-l$ arrows from $J$ to $I$ if $l\leq0$.

The mutation $\mut_K (\sfQ)$ is an involution that corresponds to a change of the quiver and
therefore to a change of basis for the lattice $\widehat{\Lambda}$ which
is reflected in a redefinition of the generators $\sfX_I$ of
$\torus_{\widehat{\Lambda}}^{*}$. As in~\cite{keller}, we neither
mutate nor draw arrows between framing vertices $\bullet$; the
elements $\sfX_\infty$ then play the role of coefficients in the
corresponding cluster algebra. For $K=r\in\widehat\Gamma$ one has
\begin{eqnarray}
\sfX_I' &:=& \mut_r(\sfX_I ) \ = \ (-\lambda)^{- \langle \gamma_I , \gamma_r \rangle \, [  \langle \gamma_I , \gamma_r  \rangle]_+ } \ \sfX_I \,  (- \sfX_r)^{[ \langle \gamma_I , \gamma_r \rangle ]_+} \ , \qquad I\neq r \ , \nonumber \\[4pt]
\sfX_r' &:=& \mut_r(\sfX_r ) \ = \ \sfX_r^{-1} \ ,
\end{eqnarray}
where $[n]_+ := \max \{ n,0 \}$ for $n\in\zed$.
The \textit{quantum} version of this operator is obtained by composing with conjugation by the quantum dilogarithm operator
\begin{equation}
\sfX_I \ \longmapsto\ {\rm Ad}_{\Psi (\sfX_r ; \lambda)}(\sfX_I):= \Psi (\sfX_r ; \lambda)^{-1} \, \sfX_I \, \Psi (\sfX_r ; \lambda) \ .
\end{equation}
Explicitly, we can write for a quantum cluster mutation at vertex $K=r$ the transformation
{ \small
\begin{equation}
\mut_r^{\lambda} (\sfX_I) = \Psi (\sfX_r ; \lambda)^{-1} \, \mut_r (\sfX_I)
\, \Psi (\sfX_r ; \lambda) = \mut_r \left(  \Psi (\sfX_r^{-1} ; \lambda)^{-1} \, \sfX_I \, \Psi (\sfX_r^{-1} ; \lambda) \right) \ .
\label{quantumut}\end{equation} }
As demonstrated in~\cite{Cecotti:2010fi}, the quantum monodromy
operator (\ref{monop}) can typically be written (up to conjugation) as
a product of quantum mutation operators (\ref{quantumut}).
Once again this operation is completely rephrased in terms of representation theory data, that is explicitly determined once the orbifold singularity (and therefore the virtual quiver) is given. 

For simplicity, from now on we neglect the framing and the D6-brane
charge in the lattice; it is straightforward to incorporate them
back. From~\cite{fock}, we know that for the quantum dilogarithm the relation
$\lambda^{-\langle \gamma_r , \gamma_s \rangle} \ \sfX_{r} \
\sfX_{s} = \lambda^{-\langle \gamma_s , \gamma_r \rangle} \ \sfX_{s} \ \sfX_{r}$ implies
\begin{equation}
\Psi (\sfX_r^{-1} ; \lambda)^{-1} \ \sfX_s = \sfX_s \ \Psi (\lambda^{2
  \langle \gamma_r , \gamma_s \rangle} \, \sfX_r^{-1} ; \lambda
)^{-1} \ .
\label{dilogid1}\end{equation}
In particular, in our case, since the pairing between fractional branes is determined by the intersection pairing on $K^c (X)$, one has
\begin{eqnarray}
\Psi  (\lambda^{2\langle \gamma_r , \gamma_s \rangle}  \, \sfX_r^{-1}
; \lambda  )^{-1} \  \Psi (\sfX_r^{-1} ; \lambda)  \hspace{-3cm} &&\\[4pt] & =&  \nonumber \Psi
(\lambda^{2a^{(2)}_{rs} - 2a^{(1)}_{rs}} \, \sfX_r^{-1} ; \lambda
)^{-1} \  \Psi (\sfX_r^{-1} ; \lambda) \label{dilogid2} \\[4pt] 
& = &
\left\{ \ 
\begin{matrix} 
\displaystyle{\prod_{i=1}^{a^{(2)}_{rs}
  - a^{(1)}_{rs}}}\ \Big( 1 + \lambda^{2i-1}\ \sfX_r^{-1}\Big) \quad & \text{if} \quad a^{(2)}_{rs}
> a^{(1)}_{rs} \ , \\[4pt]
1 \quad &  \text{if} \quad a^{(2)}_{rs} = a^{(1)}_{rs} \ , \\[4pt]
\displaystyle{\prod_{i=1}^{a^{(1)}_{rs}
  - a^{(2)}_{rs}}}\ \Big( 1 + \lambda^{1-2i} \ \sfX_r^{-1}\Big)^{-1}
\quad &  \text{if} \quad a^{(2)}_{rs} < a^{(1)}_{rs} \ .
\end{matrix} \right.\nonumber
\end{eqnarray}
Note that the non-trivial products have $\big|a^{(2)}_{rs} - a^{(1)}_{rs} \big| $ terms.

Let us consider an explicit example of the action of a quantum
mutation on our instanton quivers. For the virtual quiver
(\ref{C3Z6virtual}) of $\complex^3 / \zed_6$, let us mutate at the
vertex $V_2$ to get the quiver
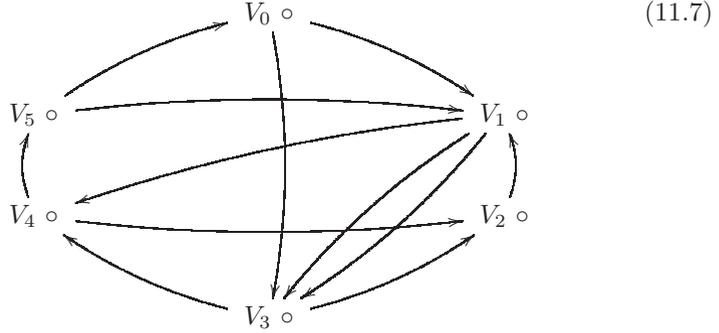
\begin{equation}
\vspace{4pt}
\begin{xy}
\xymatrix@C=20mm{
& \ V_0 \ \circ \ \ar@/^/[dr] \ar@/^0.5pc/[ddd] & \\
V_5 \ \circ \ \ar@/^/[ur]  \ar@/^/[rr] & & \ V_1 \ \circ  \ar@/^/[ddl] \ar@/_/[ddl] \ar@/_/[lld]  \\
V_4 \ \circ \ \ar@/^/[u] \ar@/_/[rr]  & &  \ V_2 \ \circ \ar@/_/[u]  \\
& \ V_3 \ \circ \ \ar@/^/[ul] \ar@/_/[ur] & 
}
\end{xy}
\vspace{4pt}
\end{equation}
The action of the classical mutation on the quantum torus generators
is given by
\begin{eqnarray}
\mut_2 (\sfX_0) &=&  \lambda^{- 1} \ \sfX_0 \ \sfX_2 \ = \ 
\sfX_{\gamma_0+\gamma_2} \ , \nonumber \\[4pt]
\mut_2 (\sfX_1) &=&  \lambda^{- 1} \ \sfX_1 \ \sfX_2 \ = \ \sfX_{\gamma_1 +
  \gamma_2} \ , \nonumber \\[4pt]
\mut_2 (\sfX_2) &=& \sfX_2^{-1} \ , \nonumber \\[4pt]
\mut_2 (\sfX_3) &=& \sfX_3 \ , \nonumber \\[4pt]
\mut_2 (\sfX_4) &=& \sfX_4 \ , \nonumber \\[4pt]
\mut_2 (\sfX_5) &=& \sfX_5 \ ,
\end{eqnarray}
where we have used the twisted multiplication rule
\begin{equation}
\sfX_{\gamma_r+\gamma_s} = \lambda^{-\langle \gamma_r , \gamma_s \rangle} \ \sfX_{r} \ \sfX_{s}
\end{equation}
to make the change of basis in the quantum torus explicit. The quantum
mutation is therefore given by
\begin{eqnarray}
\mut_2^{\lambda} (\sfX_0) &=&  \lambda^{- 1} \ \sfX_0 \ \sfX_2 \,( 1 +
\lambda \ \sfX_2 ) \ , \nonumber \\[4pt]
\mut_2^{\lambda} (\sfX_1) &=&  \lambda^{- 1} \ \sfX_1 \ \sfX_2 \,( 1 +
\lambda \ \sfX_2 ) \ , \nonumber \\[4pt]
\mut_2^{\lambda} (\sfX_2) &=& \sfX_2^{-1} \ , \nonumber \\[4pt]
\mut_2^{\lambda} (\sfX_3) &=& \sfX_3 \, ( 1 + \lambda^{- 1} \ \sfX_2 )^{-1} \ , \nonumber \\[4pt]
\mut_2^{\lambda} (\sfX_4) &=& \sfX_4  \,( 1 + \lambda^{-1} \ \sfX_2 )^{-1} \ , \nonumber \\[4pt]
\mut_2^{\lambda} (\sfX_5) &=& \sfX_5 \ .
\end{eqnarray}
This example illustrates that the rules of quantum mutations are
completely determined by the representation theory data associated
with the singularity: one can literally see the structure of the third
column of the intersection matrix (\ref{intC3Z6}) in $\mut_2^{\lambda} (\sfX_r)$. 

To understand completely the quantum cluster algebra one should
continue to compose quantum mutations. Of course, having changed
basis in the quantum torus, the intersection pairing has changed as
well and is still determined by the arrow structure of the mutated quiver. It
is shown in \cite{fock} that the new pairing $\langle-,-\rangle'$ can be obtained from the
old pairing $\langle-,-\rangle$ through
{ \scriptsize
\begin{equation}
\langle \gamma_r  , \gamma_s \rangle' =
\left\{ \rule{0cm}{1cm} \right.
 \begin{matrix}
- \langle \gamma_r , \gamma_s \rangle \ & \text{if} \quad K
\in \{r,s\} \ , \\[4pt]
\langle \gamma_r , \gamma_s \rangle \ & \text{if} \quad
\langle \gamma_r , \gamma_K \rangle \ \langle \gamma_K
, \gamma_s \rangle \le 0 \ , \quad K \notin \{r,s\} \ ,
\\[4pt]
\langle \gamma_r , \gamma_s \rangle + \big\vert \langle \gamma_r ,
\gamma_K \rangle \big\vert \ \langle \gamma_K , \gamma_s \rangle \ 
& \text{if} \quad \langle \gamma_r , \gamma_K \rangle \
\langle \gamma_K , \gamma_s \rangle > 0 \ , \quad K
\notin \{r,s\}  \ .
\end{matrix}
\end{equation} }
Again all the subsequent pairings are determined by the singularity structure and in particular the form of the quantum mutation operator is completely determined, although doing so in practice might prove to be rather challenging. In a given chamber of the K\"ahler moduli space, these compositions
should be compared with the adjoint actions ${\rm Ad}_{\sfM
  (\lambda)}(\sfX_r)$ of the corresponding quantum monodromy operators
(\ref{monop}), which can in principle be computed explicitly by using
the quantum dilogarithm identities (\ref{dilogid1})--(\ref{dilogid2}).

Note that our mutations are concerned with the cluster algebra structure. From this perspective a mutation corresponds to a change of basis in the lattice of charges. Changing basis vectors alters the possible bound states that the fractional branes can form and hence also affects the spectrum of BPS states; it is intriguing that such a transformation in the noncommutative crepant resolution chamber induces an analogous basis change in the large radius limit via the McKay correspondence. It would be interesting to understand how this formalism is related to some notion of Seiberg duality on our instanton matrix quantum mechanics, which would be a statement about representations of the quiver, analogously to the discussion of e.g.~\cite{vitoria,DWZ}; this would tie in nicely with the tilted derived equivalences between the noncommutative crepant resolution and large radius chambers described in~\cite[Section~5.9]{Cirafici:2010bd}. It would seem that the correct setup to apply Seiberg--like dualities is the one of the virtual instanton quivers. However, they do not seem to have an immediate geometric interpretation in terms of tilting objects in the derived category of coherent sheaves on $X$.

\section{Discussion}

In this paper we have taken one step further the discussion initiated in \cite{Cirafici:2010bd} and clarified some loose ends. We have also used this chance to present the material of \cite{Cirafici:2010bd} in a simple form, by introducing the basic features of stacky gauge theories without all the necessary intricacies. Indeed while the formalism relies on a quite long construction which lies at the interface of noncommutative field theory, algebraic geometry and representation theory, at the end of the day it can be simply restated as an instanton counting problem, albeit with some stringy inputs. The approach can be summarized by saying that the stacky gauge theory path integral localized onto certain configurations weighted by an instanton action. These configurations correspond to certain enumerative invariants, which depend on the instanton number $\mbf k$ and a choice of boundary condition labelled by a vector $\mbf N$, which can be computed using a framed instanton quiver. Each of the aforementioned quantities is explicitly computable and the interested reader will find plenty of examples in \cite{Cirafici:2010bd}. In this note we have investigated various ways in which this formalism can be extended, refraining from extensive computations of the invariants. For example the BPS invariants admit a combinatorial ``refinement", much as ordinary Donaldson--Thomas invariants, which is obtained by a modification of the instanton action to include spin degrees of freedom. Indeed on can take a step further and consider motivic invariants $\big[\NDT_{\mu=0} (\mbf k , \mbf N)\big] $. The stacky gauge theory setup implies that one has to work in a superselection sector determined by the boundary condition vector $\mbf N$. The resulting invariant turn out to have a quite intricate dependence on the chosen boundary conditions.

Interestingly the stacky gauge theory seems to imply the existence of another object, the (framed) virtual instanton quiver, which is obtained by endowing the lattice of fractional brane charges with the natural K--theory pairing. We interpret this virtual quiver as a limit of the instanton quiver where certain bi--fundamental fields are decoupled. We suggest that this quiver has a special status in the study of BPS invariants. Indeed once it is introduced, the wall--crossing formula as well as all the machinery of quantum cluster algebras is completely determined by representation theory data via the McKay correspondence. In particular this appears to be the correct setting to study Seiberg--like dualities at the level of the instanton quantum mechanics. Unfortunately having decoupled bi--fundamental fields has most likely taken us in some other region of the moduli space. The hope is that one could use the McKay correspondence to constrain the BPS spectrum of the virtual quiver quantum mechanics, possibly along the same lines in which the instanton sums in $\cN = 4$ super Yang--Mills on a four--dimensional ALE space are determined by the characters of the associated affine Kac--Moody algebra. We feel these speculations deserve further investigations. 

\section*{Acknowledgments}

M.C. was supported in part by the Funda\c{c}\~{a}o
para a Ci\^{e}ncia e Tecnologia (FCT/Portugal). R.J.S.
was supported in part by grant ST/G000514/1 ``String Theory Scotland'' from the UK Science and Technology
Facilities Council.

\end{document}